\documentclass[pdflatex,sn-mathphys-num]{sn-jnl}
\usepackage{graphicx}%
\usepackage{multirow}%
\usepackage{amsmath,amssymb,amsfonts}%
\usepackage{amsthm}%
\usepackage{mathrsfs}%
\usepackage[title]{appendix}%
\usepackage{xcolor}%
\usepackage{textcomp}%
\usepackage{manyfoot}%
\usepackage{booktabs}%
\usepackage{algorithm}%
\usepackage{algorithmicx}%
\usepackage{algpseudocode}%
\usepackage{listings}%

\theoremstyle{thmstyleone}%
\theoremstyle{thmstyletwo}%

\theoremstyle{thmstylethree}%

\usepackage{listings}
\begin{document}
\title[Article Title]{Learning Nonlocal Interactions in Large-Scale Heterogeneous Metamaterials through Scientific Machine Learning}

\author[1]{\fnm{Dex Doksoo} \sur{Lee}}
\equalcont{These authors contributed equally to this work.}

\author[2]{\fnm{Jehyeon} \sur{Shin}}
\equalcont{These authors contributed equally to this work.}

\author[3]{\fnm{Yue} \sur{Yu}}

\author[4]{\fnm{Sunkyu} \sur{Yu}}

\author*[5,6,7,8]{\fnm{Junsuk} \sur{Rho}}
\email{jsrho@postech.ac.kr}

\author*[1]{\fnm{Wei} \sur{Chen}}
\email{weichen@northwestern.edu}

\affil[1]{%
  \orgdiv{Department of Mechanical Engineering},
  \orgname{Northwestern University},
  \orgaddress{%
    \city{Evanston},
    \postcode{60208},
    \state{IL},
    \country{USA}}%
}

\affil[2]{%
  \orgdiv{Graduate School of Artificial Intelligence},
  \orgname{Pohang University of Science and Technology},
  \orgaddress{%
    \city{Pohang},
    \postcode{37673},
    \country{Republic of Korea}}%
}

\affil[3]{%
  \orgdiv{Department of Mathematics},
  \orgname{Lehigh University},
  \orgaddress{%
    \city{Bethlehem},
    \postcode{18015},
    \state{PA},
    \country{USA}}%
}

\affil[4]{%
  \orgdiv{Department of Electrical and Computer Engineering},
  \orgname{Seoul National University},
  \orgaddress{%
    \city{Seoul},
    \postcode{08826},
    \country{Republic of Korea}}%
}

\affil[5]{%
  \orgdiv{Department of Mechanical Engineering},
  \orgname{Pohang University of Science and Technology},
  \orgaddress{%
    \city{Pohang},
    \postcode{37673},
    \country{Republic of Korea}}%
}

\affil[6]{%
  \orgdiv{Department of Chemical Engineering},
  \orgname{Pohang University of Science and Technology},
  \orgaddress{%
    \city{Pohang},
    \postcode{37673},
    \country{Republic of Korea}}%
}

\affil[7]{%
  \orgdiv{Department of Electrical Engineering},
  \orgname{Pohang University of Science and Technology},
  \orgaddress{%
    \city{Pohang},
    \postcode{37673},
    \country{Republic of Korea}}%
}

\affil[8]{%
  \orgname{POSCO-POSTECH-RIST Convergence Research Center for Flat Optics and Metaphotonics},
  \orgaddress{%
    \city{Pohang},
    \postcode{37673},
    \country{Republic of Korea}}%
}


\abstract{In multiscale systems, nonlocal interactions among spatially distributed constituents can unlock collective, potentially emergent behavior irreducible to individual constituent responses. Beyond characterization, nonlocality in multiscale systems is increasingly viewed as a design element across fracture mechanics, wave physics, and disordered systems. Existing nonlocality theories provide rigorous descriptions and mechanistic insight, but often require prescribed analytic forms, range, symmetries, or periodicity. Machine learning offers a data-driven route to learn interactions, yet principled approaches that tie the network design to physical theory remain nascent. Here we present domain-decomposition attention neural operator (DANO), which learns configuration-dependent interactions in heterogeneous systems governed by linear field equations from field data alone. Through a case study in large-scale optical metasurfaces, we demonstrate DANO's full-field prediction under geometric heterogeneity and across drive conditions, small-to-large domain transfer, and multifunctional inverse design, with learned interactions validated through physics-based interrogation. As an AI-for-Science tool, it supports interaction discovery and scalable, customizable design of nonlocal multiscale systems, including virtual-reality optics.}

\maketitle

\tableofcontents 
\section{Introduction}

\begin{figure}[!tbp]
    \centering
    \includegraphics[width=0.95\linewidth]{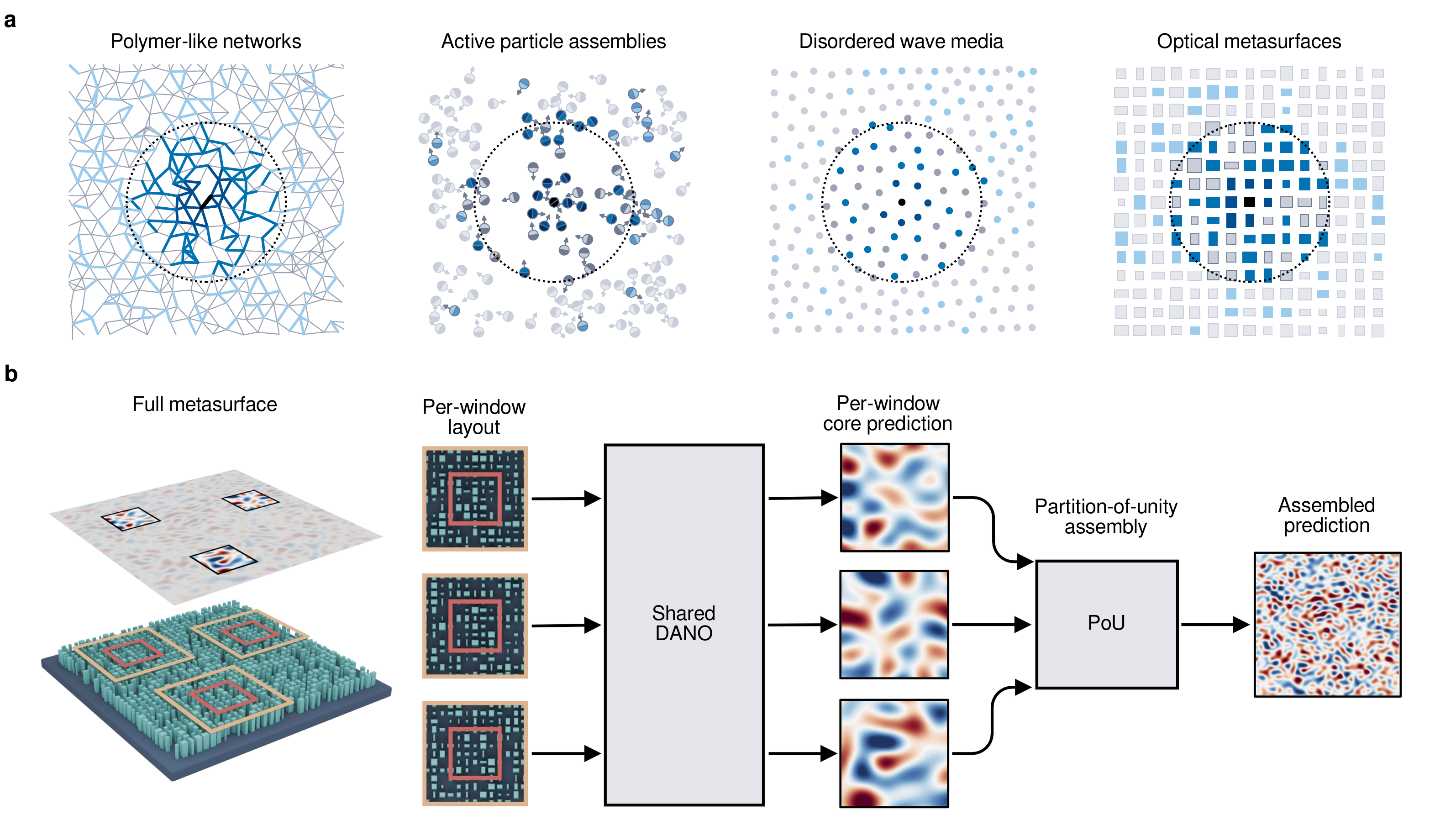}
    \caption{\textbf{Nonlocal constituent interactions and windowed prediction of a heterogeneous metasurface.} Schematics and fields are illustrative. \textbf{a}, Nonlocal interactions among distributed constituents in polymer-like networks, active particle assemblies, disordered wave media and optical metasurfaces. In each panel, the interaction of one constituent (black) with the others extends beyond its nearest neighbors (blue, darker for a stronger interaction). Dotted circles mark an interaction range, drawn with the same radius in every panel. Arrows give the propulsion directions of the active particles. \textbf{b}, Windowed prediction of the response field across a large metasurface using the domain-decomposition attention neural operator (DANO). Tan squares mark context windows, each containing an interior prediction core (red). The same DANO uses the geometry and drive within each window to predict the response field on its core. Partition-of-unity (PoU) assembly combines the core predictions into a field over the full array.}
    \label{fig:fig1}
\end{figure}

Across natural and engineered systems, collective behavior often emerges from spatially nonlocal interactions of distributed constituents that extend beyond their individual responses (Fig.~\ref{fig:fig1}a), 
leading to effectively nonlocal responses even when the governing equations are local~\cite{torquato-2021-nonlocal-effective-electromagnetic-wave-characteristics}. 
Understanding such interactions has often revealed mechanisms that are not fully attributed to the constituent-level individual responses, including nonlocal fracture energy release in polymer-like networks~\cite{deng2023nonlocal}, long-range order in disordered wave systems~\cite{yu2021engineered}, and collective resonant transport mediated by near-field interaction~\cite{li2020collective}. These interactions open a new avenue to reverse-engineer collective behaviors such as fracture resistance~\cite{meng2025dual} and light-driven collective assembly in active matter~\cite{vutukuri2020light}. Nonlocality in multiscale engineering is therefore increasingly viewed as a \textit{design} element with which to engineer, beyond characterization~\cite{wang2022nonlocal, chen2025nonlocal, zhou2026designing}.

In multiscale engineering, a diverse array of local approximations has long provided a compact-yet-expressive abstraction that bridges constituents and the assembly~\cite{hashin1983analysis, li2008introduction}. This abstraction has taken different forms across disciplines yet in general entails  \textit{separation of scale}, whereby constituent-level behavior is described through effective properties and reused in predicting the system-level behavior, thereby decoupling scales either top-down or bottom-up. In metamaterials---a special class of multiscale systems whose functionality comes mainly from geometry rather than material composition~\cite{yu2018mechanical}---this principle supports the notion of constituent libraries or ``lookup-table'' approaches~\cite{mansouree2021large,park2019allglass,wang2025golay}. Capitalizing on the notion, AI-driven design for metamaterials accelerates the system-level multiscale design through constituent-level structure-to-property surrogates and generative models for spectral inverse design~\cite{liu2018generative,so2019designing}, nonlinear properties~\cite{bastek2023inverse, zheng2026algebraic}, and multiscale structural optimization~\cite{wang2020deep, wang2022ih}, as reviewed in a broader context~\cite{zheng2023deep,lee2024data}. Locality is often harnessed in system-level design, too, where local growth rules produce irregular architected materials~\cite{liu2022growth,jia2024topology, xiang2026one} or self-organized mechanical metamaterials~\cite{fromenteze2026morphogenetic} with effective local properties spatially programmed on-demand to achieve desired system-level behavior.  

Despite covering this broad realm, separation of scale faces fundamental limitations when bringing multiscale engineering to fruition.  The decoupling with separation of scale becomes less reliable when interactions cannot be neglected~\cite{an2022deep, cai2020inverse, lin2022fast, ha2023physics}, such as higher-order effects introducing long-range interactions~\cite{matlack2018designing}. Interaction has often been treated as an undesirable effect to suppress under separation of scale~\cite{wang2018meta, she2018large}, while enforcing locality is known to possibly preclude theoretically achievable performance~\cite{chung2020high,isnard2024advancing,mansouree2021large} and to often lead to geometric designs that are challenging to fabricate, such as high-aspect-ratio pillars in nanoscale fabrication~\cite{isnard2024advancing,park2024allglass}. 
More fundamentally, locality leaves limited room for deliberately exploiting collective or emergent responses that only arise from nonlocal interactions---one of the founding scientific motivations of multiscale engineering despite its added complexity.

When nonlocal interactions are viewed as a \textit{design} element, the central challenge lies in describing how distributed interactions govern system-level response. Multiple-scattering theory describes these interactions across linear wave systems, from acoustics and quantum mechanics to electromagnetics, through the Foldy–Lax equations and their recursive expansion, known as the Born series~\cite{foldy-1945-multiple-scattering-waves-i-general,lax-1951-multiple-scattering-waves,lippmann-1950-variational-principles-scattering-processes-i}. In metasurfaces, coupled-dipole~\cite{chen-2017-general-point-dipole-theory-periodic,babicheva-2019-analytical-model-resonant-electromagnetic-dipole-quadrupole} and T-matrix models~\cite{beutel-2023-treams-t-matrix-scattering-code} apply this description to meta-atoms and explain collective lattice resonances~\cite{guo-2016-geometry-dependence-surface-lattice-resonances} and their sensitivity to disorder~\cite{zakomirnyi-2019-collective-lattice-resonances-disordered-quasi-random}. Their closed-form solutions often require periodicity, symmetry or the dipole limit, and aperiodic arrays must be solved anew for each configuration~\cite{chen-2017-general-point-dipole-theory-periodic,beutel-2023-unified-lattice-sums-accommodating-multiple}. Heterogeneous (i.e., spatially varying constituent by constituent) systems hold promise yet fall outside these assumptions. System-level numerical optimization has provided a route to designing heterogeneous metasurfaces without needing to explicitly address interactions~\cite{so2023revisiting,elsawy2020numerical,ueno2024ai, cai2020inverse, sang2022toward,xu2022emerging, isnard2024advancing, mansouree2021large, ponomareva2025torchgdm}. However, computational and memory costs incurred in such system-level simulations are much more resource-intensive than constituent-level ones and escalate rapidly as the system size grows~\cite{hughes2021perspective,zhelyeznyakov2023large,li2022inverse}. In addition, after optimization the underlying interactions remain implicit in repeated system-level simulations rather than being learned as a reusable constitutive prior that could accelerate subsequent design or help address associated scientific queries. 

With the surge of scientific machine learning (SciML)~\cite{karniadakis2021physics, li-2021-fourier-neural-operator-parametric-partial, lu2021learning}, there has been growing interest in interaction engineering from deduction to induction: from predicting responses under a prescribed nonlocal interaction law to \textit{learning} a reusable interaction law itself from data.
When interactions' form, range, and strength vary with geometry and operating conditions of the system, machine learning with relational learning capabilities provides a compelling route. 
Early work in photonics reported the use of surrogates that compensate for locality-based effective properties through convolutional neural networks (CNN)~\cite{an2022deep}, while subsequent work further explored metasurfaces with geometrically aperiodic and random constituents in large domains~\cite{majorel2022deep, xu2026deep, kanmaz2023deep}. However, conventional CNNs retain a strong locality bias, so their vanilla implementations do not explicitly represent the distributed interactions dictating strongly nonlocal systems. 
Graph and attention models provide more explicit relational representations of constituent interactions~\cite{khoram2023graph,bao2025gat, wang2026novel}, while broader SciML approaches introduce physics through partial differential equation (PDE) residuals, reduced-order models, or operator learning~\cite{zhelyeznyakov2023large,lupoiu2025multi, zhelyeznyakov2021deep, augenstein2023neural}. 
Despite these advances, principled approaches that tie learned distributed nonlocal interactions to physical theory remain elusive, limiting their use for scientific interrogation and scalable nonlocal multiscale design. 
These challenges become particularly acute at large system sizes that commercialization of these multiscale systems demand~\cite{kim2023scalable,hoang2026roll,park2019allglass,park2024allglass,hou2024single,froch2025beating}, where both the number of constituents and the physical range of their interactions are of practical concern.

In multiscale systems formed by heterogeneous constituents, interactions are generally configuration-dependent. For wave propagation systems governed by linear field equations, this dependence is structured: changing the configuration changes the response operator, whereas at a fixed configuration the operator acts linearly on the incident field. For the learning problem considered here, an ideal framework would therefore respect this intrinsic structure.
Within operator learning~\cite{li-2021-fourier-neural-operator-parametric-partial,lu2021learning, kovachki2023neural}, neural Green's operators enforce this linearity by mapping the coefficient field to a Green's operator that acts linearly on the source~\cite{melchers-2026-neural-greens-operators-parametric-partial}. Their formulation is expressed at the field/operator level rather than explicitly through interactions among discrete constituents. In related work, nonlocal attention operators interpret attention as a data-dependent interaction kernel for learning operators from observed input--output pairs~\cite{yu2024nonlocal}. Together, these developments motivate a formulation tailored to large-scale heterogeneous multiscale systems, in which configuration-dependent interactions are represented explicitly among constituents while spatially resolved prediction is retained beyond the training-domain size to enable small-to-large domain transfer.

In this work, we propose the domain-decomposition attention neural operator (DANO) to learn this family of configuration-dependent response operators (Fig.~\ref{fig:fig1}b). Motivated by self-consistent multiple-scattering theory, a geometry path contextualizes the constituent geometries and determines the coefficients of a separate field path. The field path is complex-linear in the drive (i.e., the externally applied excitation) by construction, while the dependence of its coefficients on the configuration is learned from numerical solutions of the governing equations. DANO reuses the same model across overlapping windows with surrounding context and assembles their predictions into a full-domain field. Through a case study in large-scale heterogeneous (geometrically aperiodic) optical metasurfaces with varying incidence angles, we evaluate prediction on unseen configurations of constituents and its transfer across drive conditions, spatial discretization and system domain size. Upon the statistical validations, we then probe the learned operator through changes in the drive and constituent geometry, revealing signatures of propagation and configuration-dependent interaction range as physics-based interrogation. Using the same model, we also report the inverse design results of an angle-multiplexed hologram generated by a metasurface over an order of magnitude larger than the training domain, demonstrating small-to-large domain transfer.

\section{Results}\label{sec2}

\subsection{Nonlocal interactions among discrete constituents}

Consider a system of spatially distributed constituents whose field solution $u$ satisfies a local linear partial differential equation (PDE), $L[a(x)]u(x)=f(x)$. Here $L[\cdot]$ denotes the linear operator containing derivatives, the coefficient $a$ represents the constituent configuration---such as a set of material and structural parameters---and $f$ is an external drive. Because $L$ contains spatial derivatives, a field produced by one constituent propagates beyond it and can change the field exciting another constituent and, consequently, its induced response. We refer to this cross-constituent influence as an \textit{interaction}. Both these interactions and field propagation contribute to nonlocal responses that extend across space. We use \textit{nonlocality} to describe the spatial range of a specified response arising from the solution of the local PDE; this differs from a nonlocal governing equation in which spatial coupling is prescribed directly in the form of $\tilde a(x,x')$. The problem is to determine how such responses depend on the configuration and the applied drive.

We examine the system through two controlled changes: varying the drive while holding the configuration fixed, and conversely varying the configuration at fixed drive. They define
\begin{equation}
  G[a](x,x') =\left.\frac{\delta u(x)}{\delta f(x')}\right|_a,
  \qquad
  J[a;f](x,x') =\left.\frac{\delta u(x)}{\delta a(x')}\right|_f .
  \label{eq:two_kernels}
\end{equation}
The first is the source-to-solution Green’s kernel: it describes how a configured system responds to a change in its drive. The second is the field sensitivity to a local coefficient change under a specified drive: it describes how changing a constituent affects the solution in its surrounding configuration. A coefficient change perturbs $L$ by $\delta L$ and enters as an equivalent source $-(\delta L)\,u$, so its effect propagates through the same configuration-dependent Green operator $G[a]$. These two directions let us examine how the system responds both to external driving and to changes in its constituents.

Choose a background coefficient $a_0$, and denote the constituent-induced change by $\delta a=a-a_0$. The solution can then be expressed in the form of multiple scattering:
\begin{equation}
  u \;=\; u_b + \mathcal{T}[\delta a]\,u_b,
  \qquad
  \mathcal{T}[\delta a] \;=\; G_0\,V\,\bigl(I-G_0V\bigr)^{-1},
  \label{eq:induced_form}
\end{equation} where $u_b=G_0f$ is the background solution, $G_0=L[a_0]^{-1}$, and $V=L[a_0]-L[a]$ is supported where the constituents change the background operator. The term $G_0V$ gives the first-order contribution in $V$, while the inverse accounts for repeated interactions with the constituents~\cite{lippmann-1950-variational-principles-scattering-processes-i,foldy-1945-multiple-scattering-waves-i-general,lax-1951-multiple-scattering-waves}. Thus, the constituent-induced response is linear in $u_b$ at fixed configuration and generally nonlinear in $\delta a$, as examined for nonlinear expressivity in linear-optics computing~\cite{xia-2024-nonlinear-optical-encoding-enabled-by}. We learn this family of configuration-dependent response operators from full-system solutions, then use it for field prediction and for the two response measurements defined above. In the next section, we specialize the problem to electromagnetic scattering from heterogeneous metasurfaces.

\subsection{Nonlocal interactions in heterogeneous metasurfaces}

We instantiate the preceding nonlocal interaction formalism in linear electromagnetic scattering, where the constituent configuration (the spatial arrangement and geometry of constituents) determines a scattering operator and the incident field provides the drive on which it acts. This separation provides a controlled setting in which geometric variation samples a family of configuration-dependent operators, while illumination variation spans the space of incident fields used to interrogate their responses. We consider a finite $33\times33$ array of silicon nitride (SiN) pillars with rectangular cross-sections on a silica substrate, illuminated at a wavelength of 532\,nm (Fig.~\ref{fig:fig1}b). The simple rectangular geometry keeps the constituent-level response interpretable. To support generalization to unseen configurations, we systematically design the arrays so that the training data contain many distinct interaction environments. At each lattice site, the pillar cross-section is independently and uniformly selected from a prescribed library (Methods, ‘Physical system and learning target’), and the site is left vacant with probability 0.2. Size disorder and vacancies thus let the dataset sample a broad family of configuration-dependent operators. We refer to one realized array as a layout, denoted $g$. In Maxwell's equations, the layout enters through the permittivity contrast $\Delta\varepsilon_g(\mathbf{r})=\varepsilon(\mathbf{r})-\varepsilon_{\mathrm{bg}}(\mathbf{r})$, where $\varepsilon_{\mathrm{bg}}(\mathbf{r})$ is the permittivity of the bare substrate, the planar air--silica background without pillars. This contrast plays the role of $\delta a$. The background field $\mathbf{E}_b$, the same illumination on the bare substrate, serves as the drive. The scattered field $\mathbf{E}_s=\mathbf{E}-\mathbf{E}_b$ is the constituent-induced response.

The collective response of these arrays is not determined by responses assigned to each constituent independently of its surroundings. For each layout, a finite-difference time-domain (FDTD) simulation provides the reference solution, which we treat as the ground truth throughout this work. The learning target is the scattered field recorded on a 350-nm plane above the tops of the pillars, defined relative to the same illumination on the bare substrate (Extended Data Fig.~\ref{ed:data}). A local phase approximation, an instantiation of separation of scale that assigns each pillar the response of its corresponding periodic unit cell, fails to reproduce the ground-truth response: on a fully populated random array, the predicted phase differs from the ground-truth phase by 1.45\,rad root-mean-square (Extended Data Fig.~\ref{ed:lpa}). The same approximation has also been reported to fail in high-numerical-aperture metalenses, in which adjacent constituents (meta-atoms) vary rapidly~\cite{mansouree2021large,lin2022fast}. Identical pillar cross-sections appearing at different sites exhibit different responses depending on their surroundings. The collective field therefore depends on the distributed configuration rather than on transferable constituent responses alone.

Geometry and illumination play distinct roles in the dataset: geometric variation samples configuration-dependent operators, whereas illumination variation probes the incident-field space on which they act. We construct a plane-wave corpus of 426 arrays across 33 incidence directions and a structured-illumination corpus of 3,600 arrays driven by 300 band-limited random-phase speckle fields (Methods, ‘Illumination corpora and data partitions’). The structured illumination is introduced to span a broader subspace of the incident-field space rather than simply to increase the number of samples. Because scattering is linear in the incident field, the dimension of this subspace is determined by the number of independent illumination directions represented in the corpus. Window-level analysis shows that speckle illumination spans this subspace far more efficiently than plane waves (Extended Data Fig.~\ref{ed:data}e).

\subsection{A physics-structured neural operator for nonlocal interactions}

\begin{figure}[!tbp]
    \centering
    \includegraphics[width=0.95\linewidth]{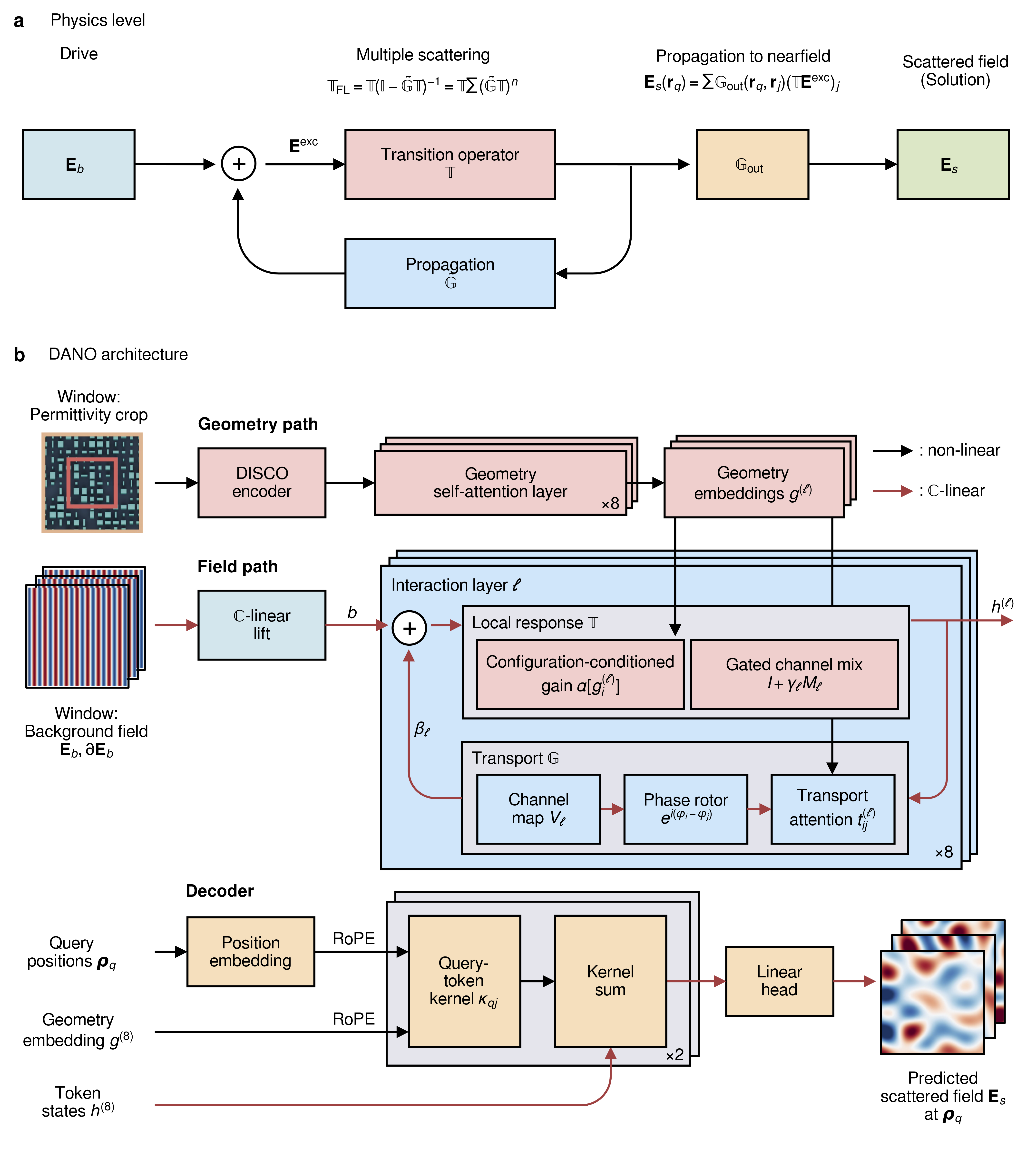}
    \caption{\textbf{Field-mediated interactions among constituents and their representation in DANO.}
    Box colors connect the physical operations in \textbf{a} with their learned counterparts in \textbf{b}: light blue, the drive and propagation between constituents; red, constituent response and its configuration-conditioned representation; orange, mapping to observation points; green, the scattered field. \textbf{a}, In electromagnetic scattering, the background field $\mathbf{E}_b$ and fields propagated from other constituents form the exciting field $\mathbf{E}^{\mathrm{exc}}$. Each constituent responds through $\mathbb{T}$, and $\mathbb{G}_{\mathrm{out}}$ maps those responses to the scattered field $\mathbf{E}_s$. \textbf{b}, For one prediction window, the geometry path encodes the constituent configuration independently of the drive. The field path repeatedly transports states between constituents, reinjects the drive and applies a configuration-conditioned response. A kernel-sum decoder maps the resulting states to the scattered field at observation points. Red arrows indicate operations that are complex-linear in the field. $\times$8 and $\times$2, numbers of stacked layers.}
    \label{fig:fig2}
\end{figure}

DANO (domain-decomposition attention neural operator) is built around the self-consistent interactions of linear field-mediated systems. The response can be expressed in Lippmann–Schwinger form~\cite{lippmann-1950-variational-principles-scattering-processes-i}, as in Eq.~\eqref{eq:induced_form}. For spatially localized constituents, grouping the induced responses at the constituent level yields a Foldy–Lax-type relation~\cite{foldy-1945-multiple-scattering-waves-i-general,lax-1951-multiple-scattering-waves}: each constituent responds to a field shaped by the responses of its surroundings (Fig.~\ref{fig:fig2}a). For the metasurface considered here, this relation takes the form
\begin{equation}
  \mathbf{E}_s \;=\; \mathbb{G}_{\mathrm{out}}\,\mathbb{T}\,\bigl(\mathbb{I}-\tilde{\mathbb{G}}\mathbb{T}\bigr)^{-1}\mathbf{E}_b \;\equiv\; \mathcal{T}[g]\,\mathbf{E}_b,
  \label{eq:result_master}
\end{equation} where $\mathbb{T}$ collects the transition operators of individual constituents, $\tilde{\mathbb{G}}$ propagates responses between constituents, and $\mathbb{G}_{\mathrm{out}}$ maps them to the observation plane. The inverse accounts for repeated interactions. The whole-array scattering operator $\mathcal{T}[g]$ is the metasurface form of $\mathcal{T}[\delta a]$ in Eq.~\eqref{eq:induced_form}, restricted to the observation plane: it maps the background field to the constituent-induced field. It depends nonlinearly on the layout $g$, while its action on $\mathbf{E}_b$ is linear.

DANO follows this multiple-scattering structure with separate geometry and field paths (Fig.~\ref{fig:fig2}b). In the geometry path, each occupied or vacant site is initially encoded from its local constituent geometry. Discrete–continuous (DISCO) convolutions construct these local descriptors using kernels defined in physical coordinates, allowing evaluation on different spatial meshes (Extended Data Fig.~\ref{ed:invariants}a; Methods, ‘Geometry representation’). Self-attention then incorporates the surrounding configuration into the site representations. The geometry path does not receive the drive, so it determines the learned interaction operator independently of the drive.

As an analog of the repeated interactions encoded by the inverse in Eq.~\eqref{eq:result_master}, which corresponds to multiple orders of scattering in its Born series, the field path uses an unrolled recurrence. Each update transports field states between constituents, reinjects the drive and applies configuration-conditioned local responses (Methods, ‘Linear field path’). A kernel-sum decoder maps the resulting states to the scattered field at query positions $\boldsymbol{\rho}_q$ on the observation plane (Methods, ‘Decoder and full-field assembly’). All operations acting on the field are complex-linear, with coefficients determined by the geometry path; DANO is therefore exactly complex-linear in the drive by construction.

To extend predictions to large arrays, DANO applies the learned interaction operator to spatial windows with a surrounding context halo (Fig.~\ref{fig:fig1}b). Each window predicts the field on an interior core; the halo supplies surrounding geometry and field context without prescribing a hard physical interaction range (Methods, ‘Windows and context halo’). The chosen halo width follows the validation study in Extended Data Fig.~\ref{ed:window}c. Overlapping core predictions are combined by linear partition-of-unity assembly, allowing independently evaluated windows to form a full-domain field (Methods, ‘Decoder and full-field assembly’). The following section tests prediction accuracy when this operator is applied to new illuminations, meshes, layouts and array sizes.
\subsection{Predictive accuracy and transfer}

\begin{figure}[!tbp]
    \centering
    \includegraphics[width=0.95\linewidth]{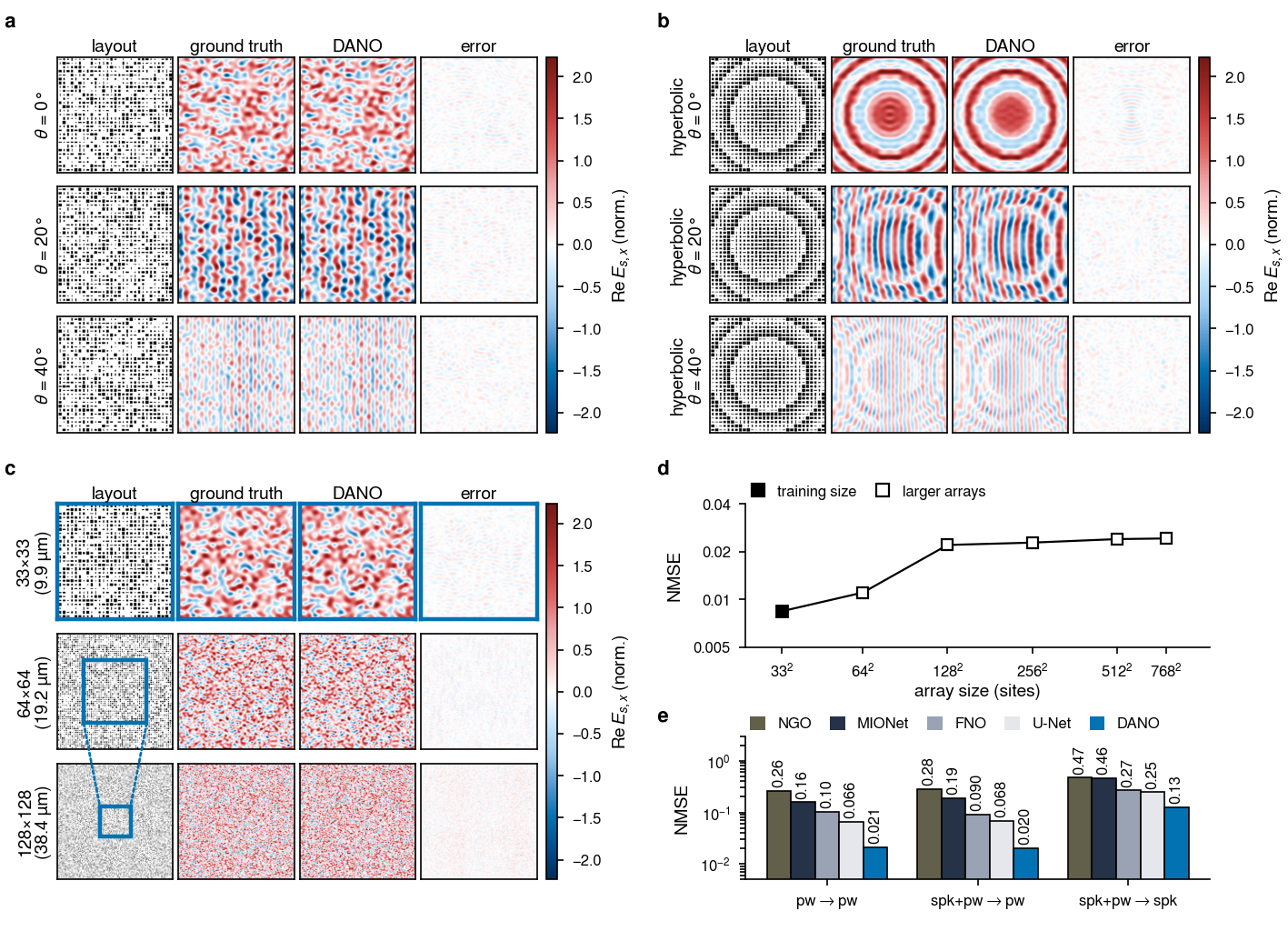}
    \caption{\textbf{Full-field prediction accuracy across configurations and domain sizes.} 
    \textbf{a}, In-distribution prediction on held-out random layouts under plane-wave illumination at $\theta=0^\circ$, $20^\circ$ and $40^\circ$ (rows). Columns show the layout (black, SiN pillars), the ground truth from finite-difference time-domain simulation, the domain-decomposition attention neural operator (DANO) prediction and their error (DANO minus ground truth) for Re$\,E_{s,x}$, normalized by the root-mean-square incident-field amplitude. \textbf{b}, Structured-layout transfer to a $33\times33$ hyperbolic metalens with a 10-\textmu m focal length, from a layout family absent from training. Rows show the same incidence angles, and columns follow \textbf{a}. \textbf{c}, Domain-size transfer from the $33\times33$-site training size to $64\times64$ and $128\times128$ sites at normal incidence, using one independently drawn layout at each size. The blue frame marks the training-array size; blue squares connected by dotted guides mark its 9.9-\textmu m width for scale. \textbf{d}, Normalized mean squared error (NMSE) of the assembled field against the ground truth for one layout at each array size. \textbf{e}, Mean test NMSE of DANO and the neural Green's operator (NGO), multiple-input operator network (MIONet), Fourier neural operator (FNO) and U-Net baselines, all trained on the same windows. Group labels give training corpus $\rightarrow$ test illumination, with pw for plane waves and spk for speckle. Each bar is one training run evaluated on 213 plane-wave or 600 speckle test cases.}
    \label{fig:fig3}
\end{figure}

We evaluated field prediction on a separately generated test set of random layouts, followed by transfer tests involving a new illumination family, finer spatial meshes, a structured layout and larger arrays. For each case, we calculated the normalized mean squared error (NMSE) of the scattered field against the ground truth using both transverse field components. The random-layout test set comprised 213 plane-wave and 600 speckle cases with layouts absent from training and validation (Methods, ‘Evaluation protocol’). For each training-corpus setting, the neural Green's operator (NGO), MIONet, the Fourier neural operator (FNO) and U-Net were trained on the same window draws and evaluated on the same test cases as DANO (Methods, ‘Benchmarking’). All subsequent transfer tests used the DANO model trained on both illumination corpora without retraining.

On the held-out random layouts, DANO reproduced the ground-truth fields across the tested incidence angles (Fig.~\ref{fig:fig3}a). The model trained on both corpora achieved mean NMSEs of 0.0200 for plane waves and 0.1269 for speckle, compared with 0.068 and 0.25, respectively, for the best-performing baseline, U-Net (Fig.~\ref{fig:fig3}e). Across five training seeds, the corresponding errors were 0.0201 $\pm$ 0.0003 and 0.1295 $\pm$ 0.0023 for DANO and 0.0666 $\pm$ 0.0007 and 0.2463 $\pm$ 0.0010 for U-Net (mean $\pm$ s.d.). A variant that keeps the transport step but restricts it to each constituent's own state had test errors about 2.1 times higher for both illuminations. A fully local variant, in which each constituent responds independently of its neighbors, had errors 11 and 3.5 times higher for plane waves and speckle.

A DANO model trained only on plane waves reached 0.021 on the plane-wave test, close to the 0.0200 obtained after training on both corpora (Fig.~\ref{fig:fig3}e). Compared with the discrete plane-wave training directions, speckle fields combine angular components in many ways and span a larger window-level illumination space (Extended Data Fig.~\ref{ed:data}c,e). At fixed layout, both the scattering response and DANO are complex-linear in the drive (Methods, ‘Physical system and learning target’ and ‘Linear field path’). Speckle training therefore probes the configuration-dependent response under a richer set of drives, consistent with the preserved plane-wave accuracy.

The speckle error nevertheless remains about six times the plane-wave error. Angular components that propagate in the silica substrate but not in air account for 74\% of this error (Extended Data Fig.~\ref{ed:speckleband}a,b,d; Methods, ‘Source-band decomposition’). Without these components, the mean NMSE is 0.040, close to the 0.039 obtained for plane waves at 40\(^\circ\) (Extended Data Fig.~\ref{ed:speckleband}c).

Gaussian-beam illumination, another family of drives unseen during training, tests this drive-space transfer. A Gaussian beam is a coherent superposition of plane waves, and DANO’s prediction obeys the same superposition rule because its drive linearity is built into the architecture rather than imposed through training (Methods, ‘Mesh transfer and linearity tests’). The dominant angular components of these beams lie within the range represented by the speckle training fields (Extended Data Figs.~\ref{ed:data}c and~\ref{ed:gaussian}b). On one held-out layout, DANO reproduced the ground-truth field patterns for four normally incident beams with waists from 0.5 to 2\,\textmu m and had lower prediction error than each baseline at every waist (Extended Data Fig.~\ref{ed:gaussian}a,c). This result supports transfer to new combinations of angular components within the sampled spectral range.

We also examined whether the learned model transfers across spatial discretizations. Without retraining, DANO reproduced the main ground-truth field patterns on 12.5-nm and 9.375-nm meshes, finer than the 20-nm training mesh (Extended Data Fig.~\ref{ed:invariants}b). The learned kernels were evaluated on the new grids while their parameters and physical window dimensions remained fixed (Methods, ‘Mesh transfer and linearity tests’). Across 50 layouts, mean NMSE increased from approximately 0.01 on the training mesh to approximately 0.03 on the finest mesh (Extended Data Fig.~\ref{ed:invariants}d). The model thus transferred to the finer meshes, with a measurable loss of accuracy.

In addition to validation on random layouts, we tested the DANO model on a deliberately structured array absent from training to assess transfer beyond the random-layout distribution. For this purpose, we considered a 33\(\times\)33 hyperbolic metalens~\cite{khorasaninejad2016metalenses} designed with the local phase approximation (Fig.~\ref{fig:fig3}b). At incidence angles of 0\(^\circ\), 20\(^\circ\) and 40\(^\circ\), DANO reproduced the main spatial features of the ground-truth scattered fields. This one-device comparison supports compositional generalization at the array level within the training domain size.

As another transfer test, we examined the fully ordered limit of a structured layout: a uniform array, which repeats one pillar cross-section on every site and is absent from training. We evaluated eight such 33\(\times\)33 arrays under normally incident plane waves (Extended Data Fig.~\ref{ed:ordered}a). The NMSE ranged from 0.038 to 0.99, compared with a test mean of 0.0096 at normal incidence on random layouts (Extended Data Fig.~\ref{ed:ordered}b). One complex factor per array, fitted to the ground truth, removed 81–99\% of the squared error and left NMSEs of 0.005–0.047, below 0.0096 for five of the eight arrays. In the ordered limit, DANO thus retained the overall spatial field pattern, and its error was dominated by a global amplitude and phase mismatch.

We finally examined whether the same DANO model transfers beyond the 33\(\times\)33 training domain. Partition-of-unity assembly joins predictions from overlapping windows (Methods, ‘Decoder and full-field assembly’); on a training-sized array, the assembled field follows the ground truth across the overlap between neighboring windows (Extended Data Fig.~\ref{ed:window}d–f). DANO also reproduced the field structure of larger arrays shown in Fig.~\ref{fig:fig3}c. For one layout at each size, assembled-field NMSE rose from approximately 0.008 at 33\(\times\)33 sites to approximately 0.02 at 128\(\times\)128 sites, then remained below 0.03 through 768\(\times\)768 sites (Fig.~\ref{fig:fig3}d). These comparisons establish size transfer over the tested layouts, while showing that accuracy is not independent of array size.

\subsection{Physics-based interrogation of learned nonlocal interactions}

\begin{figure}[!tbp]
    \centering
    \includegraphics[width=1\linewidth]{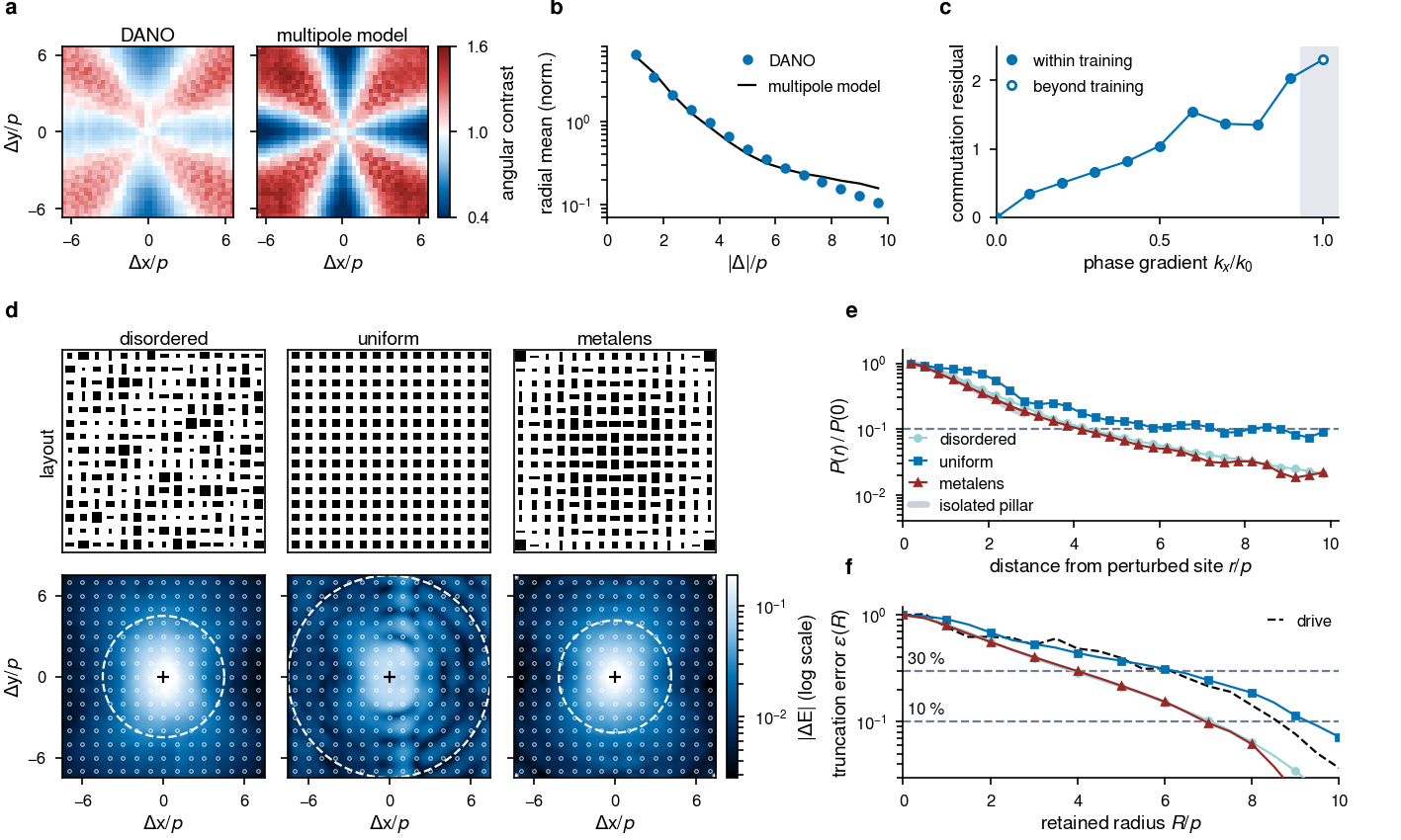}
    \caption{\textbf{Physics-based interrogation of the learned response operator.} Panels \textbf{a}--\textbf{c} vary the drive at fixed layout. Panels \textbf{d}--\textbf{f} change both side lengths of one pillar by $\pm20$\,nm at fixed drive. Distances in \textbf{a}, \textbf{b} and \textbf{d}--\textbf{f} are in units of the constituent spacing $p=300$\,nm. \textbf{a}, Angular contrast of the cross-polarized ($x\rightarrow y$) response to a localized diagnostic drive, defined as response magnitude divided by its angular mean at each radius. Left, DANO; right, a multipole model of pillar radiation. Maps are means over 71 validation windows. \textbf{b}, Radial decay of the co-polarized ($x\rightarrow x$) response to the same drive. Blue circles and black line show the DANO and multipole-model medians across the same 71 windows, respectively. \textbf{c}, Commutation residual of the learned drive-to-field operator with the phase-ramp operator $M_k$, which multiplies a field by $e^{ik_xx}$. The free-space wavenumber is $k_0$. Shading, beyond the plane-wave training coverage. \textbf{d}, Pillar footprints (top) and configuration-response magnitudes (bottom, common logarithmic scale) for disordered, uniform and metalens layouts at normal incidence. Responses are averaged over 361, 361 and 233 perturbed sites, respectively. Crosses mark perturbed sites; small circles, lattice sites; dashed circles, the 10\% radii in \textbf{e}. \textbf{e}, Radial profiles of \textbf{d}, normalized at the perturbed site, with an isolated pillar on the bare substrate as a reference. Dashed line, 10\%. \textbf{f}, Truncation error $\epsilon(R)$, the relative $L^2$ norm of the response omitted beyond radius $R$. Black dashed curve, drive response. Gray dashed lines, 30\% and 10\%.}
    \label{fig:fig4}
\end{figure}

Having established predictive accuracy, we next examine how DANO represents interactions among constituents and propagation through the background. The two response directions in Eq.~\eqref{eq:two_kernels}, the drive response $G[a]$ at fixed configuration and the configuration response $J[a;f]$ at fixed drive, organize this analysis. We probe their counterparts in the learned scattering operator $\mathcal T_{\mathbf w}[g]$, the trained approximation of $\mathcal T[g]$ in Eq.~\eqref{eq:result_master}, using localized diagnostic drives and finite changes in single constituents (Fig.~\ref{fig:fig4}; Methods, ‘Physics-based interrogation of the learned operator’).

To test whether the learned drive response propagates like physical radiation, we apply a drive concentrated around one site. DANO’s exact drive linearity allows its fixed-layout window operator to be extracted and probed directly. Across 71 validation windows, the response shares the four-lobed angular pattern and follows the radial decay of a multipole model of pillar radiation (Fig.~\ref{fig:fig4}a,b; Methods, ‘Localized drive probes and radiation reference’). The learned drive response is therefore consistent with propagation through the background, which mediates interactions between constituents.

We next ask whether the learned drive response, the counterpart of $G[a]$ in Eq.~\eqref{eq:two_kernels}, is pointwise, that is, whether the output at each position depends only on the drive at the same position. We multiply the drive by a phase that varies linearly across the array, $M_k=e^{ik_xx}$. A pointwise response would return the original output multiplied by the same phase:
\begin{equation}
  \mathcal T_{\mathbf w}[g](M_k\mathbf{E}_b)=M_k\,\mathcal T_{\mathbf w}[g](\mathbf{E}_b).
  \label{eq:commutation_metric}
\end{equation}
For one fixed window, the commutation residual, the relative mismatch between the two sides, reaches $38\%$ at the smallest tested phase gradient and grows with the gradient (Fig.~\ref{fig:fig4}c; Methods, ‘Phase-ramp commutation test’). The learned response is therefore exactly linear in the drive yet nonlocal: the output at one position depends on the drive at other positions.

Changing one constituent at fixed drive probes the configuration response $J[a;f]$: how far it extends and how this extent depends on the surrounding layout. We change both side lengths of one pillar by $\pm$20\,nm while holding the drive and background fixed. The unchanged background field makes the scattered-field difference a finite-change counterpart of $J[a;f]$ (Methods, Eq. \eqref{eq:m-configuration-response}). We denote by $P(r)$ the mean response magnitude at distance $r$ from the changed pillar. The uniform layout produces a more persistent concentric response: $P(r)/P(0)$ falls below $10\%$ after roughly seven constituent spacings, compared with four to five in the disordered and metalens layouts (Fig.~\ref{fig:fig4}d,e). Beyond this distance, the uniform profile stays near $10\%$ up to ten spacings. Although the metalens is statistically inhomogeneous and the disordered layout is statistically homogeneous, their responses have a similar extent, consistent with the accurate metalens transfer (Fig.~\ref{fig:fig3}b). The radial profiles in Fig.~\ref{fig:fig4}e give the response at each distance, but not how much of the total response lies beyond a given radius.

At a chosen tolerance for neglected response, the learned configuration response has a finite effective range within the sampled region. The truncation error $\epsilon(R)$, the relative norm of the response omitted beyond radius $R$, falls below $30\%$ after retaining 4 to 4.5 constituent spacings and below $10\%$ after 7.5 to 8.5 in the disordered and metalens layouts (Fig.~\ref{fig:fig4}f; Methods, Eq.~\eqref{eq:m-finite-response-tail}). The $10\%$ radii exceed the distances at which the radial profiles in Fig.~\ref{fig:fig4}e fall below $10\%$ because weak outer responses accumulate, and the uniform layout requires the largest radius, beyond the sampled ten spacings. The approximately 3-\textmu m measurement motivates finite-context prediction without setting a global cutoff. The halo-width and size-transfer tests assess whether DANO predicts accurately with finite windows (Extended Data Fig.~\ref{ed:window}c; Fig.~\ref{fig:fig3}d). This windowed evaluation is used for the large-array geometric inverse design that follows.

\subsection{Large-scale inverse design and scaling of full-field evaluation}

\begin{figure}[!tbp]
    \centering
    \includegraphics[width=0.92\linewidth]{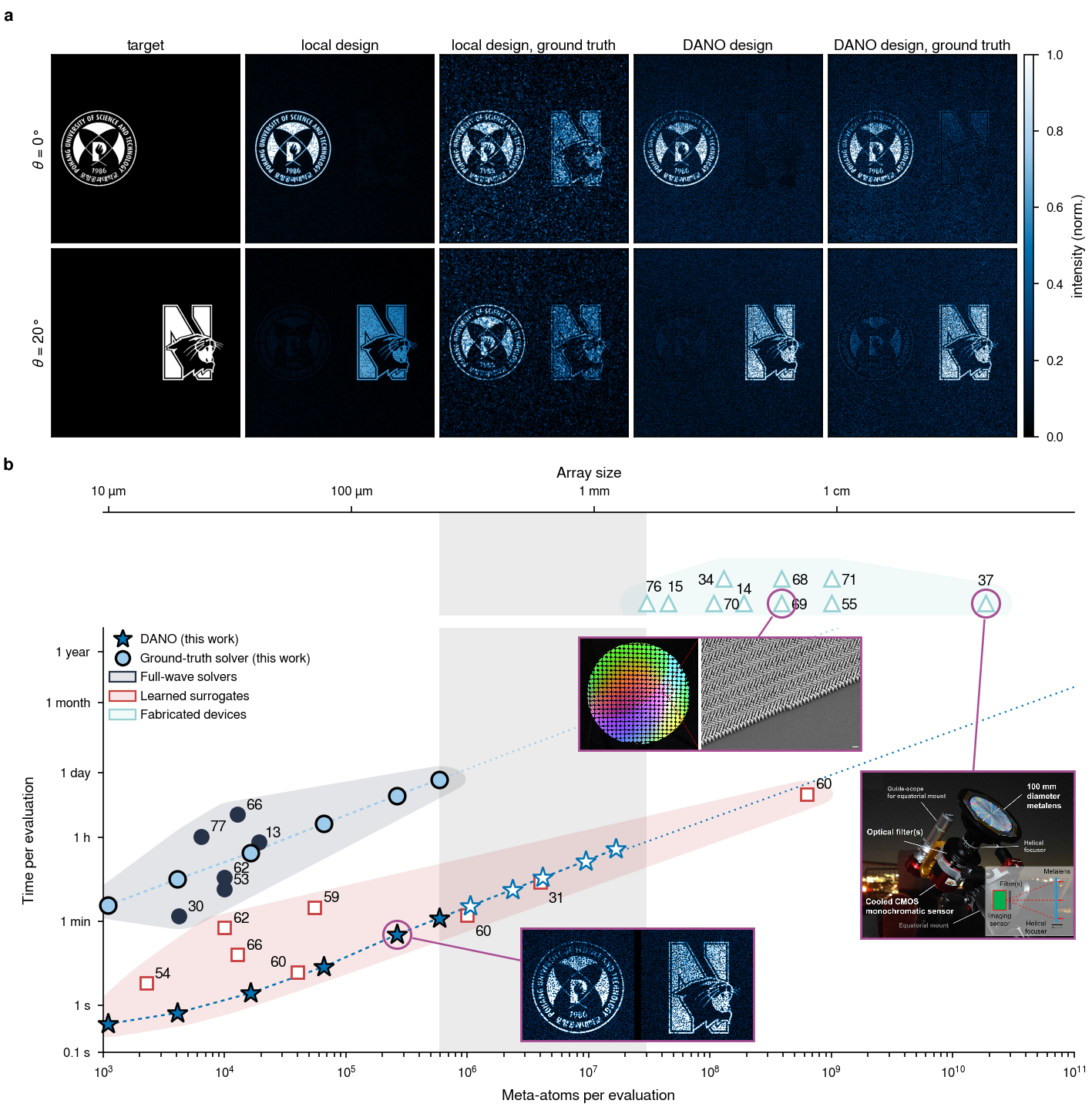}
    \caption{\textbf{Large-scale inverse design and scaling of full-field evaluation.} \textbf{a}, Far-field intensity from two $512\times512$-site layouts (153.6\,\textmu m) designed to display the POSTECH seal at $\theta=0^\circ$ and the Northwestern logo at $20^\circ$ (rows). The displayed region follows the zero order at each angle. Columns show the target, the local design evaluated by the local model and by the ground-truth solver (finite-difference time-domain simulation), and the domain-decomposition attention neural operator (DANO) design evaluated by DANO and the ground-truth solver. The local model assigns each site the angle-dependent transmission of a periodic unit cell. Each column is normalized independently. \textbf{b}, Time per evaluation versus the number of meta-atoms. Open stars show DANO times on larger arrays not simulated with the ground-truth solver. Ground-truth solver and DANO times were measured on a 64-thread node and one GPU, respectively. Dark filled circles and open squares mark full-wave solvers and learned surrogates in the literature, and open triangles mark the sizes of fabricated devices, with reference numbers beside them. Dashed lines connect the times from this work, and dotted lines extrapolate power-law fits to them. The gray band spans the interval between the largest array with whole-array ground truth and the smallest fabricated device. The top axis gives the side length of a square array at a 300-nm pitch. Insets show the hologram in \textbf{a} and two fabricated devices: a 300-mm wafer-scale metalens array with a scanning electron micrograph of its nanostructures (ref.~\citenum{hoang2026roll}; scale bar, 1\,\textmu m) and an astronomical imager built around a 100-mm metalens (ref.~\citenum{park2024allglass}). Insets in \textbf{b} reproduced from refs.~\citenum{hoang2026roll} and~\citenum{park2024allglass}, with permission from Springer Nature and under a CC BY 4.0 license, respectively.}
    \label{fig:fig5}
\end{figure}
\subsubsection{Two-drive design: angle-multiplexed hologram}\label{inv design}

We used DANO to design one fixed configuration that produces different target images for two drives: a $512\times512$-site layout that forms the POSTECH seal at normal incidence ($0^\circ$) and the Northwestern logo at $20^\circ$ (Fig.~\ref{fig:fig5}a; Methods, ‘Design task and parameterization’). Both images appear in the same output region, measured relative to the illumination direction, so the switch must come from how the array's response changes with angle. In a heterogeneous array, this angle dependence is shaped by each pillar's actual neighbors. DANO represents these neighbors, whereas a local model assumes that each pillar is surrounded by identical copies of itself. For the completed design, the ground-truth solver gave normalized cross-correlations of 0.65 and 0.71 with the intended images, compared with 0.04 and 0.08 with the other target at the respective angles. The optimized layout therefore preserves image separation.

A design based on an angle-resolved local model (Methods, ‘The inverse design problem’) loses that separation when its optimized layout is evaluated by the ground-truth solver. At $20^\circ$, the local model predicted a correlation of 0.94 with the intended Northwestern logo and 0.04 with the other target, the POSTECH seal; the solver instead gave 0.35 with the logo and 0.63 with the seal (Fig.~\ref{fig:fig5}a). The unintended image is therefore more prominent than the intended one.

To compare the two forward models directly, we used DANO to evaluate the unchanged local-design layout. DANO predicted intended-image correlations of 0.64 and 0.37 at the two angles, close to the ground-truth values of 0.65 and 0.35. At $20^\circ$, it also predicted a correlation of 0.64 with the seal, against 0.63 from the ground-truth solver. On the same layout, DANO therefore anticipates the loss of image separation that the local model misses.

\subsubsection{Evaluation time and fabricated device sizes}\label{inv design2}
Windowed inference makes full-field evaluations tractable on large arrays. For the largest array used for direct field comparisons ($768\times768$ sites), DANO evaluated the full field in about one minute on a single GPU, whereas the FDTD ground-truth solver took about 17\,h on a 64-thread node (Fig.~\ref{fig:fig5}b). DANO’s assembled-field NMSE against the ground truth remained below 0.03 for this layout (Fig.~\ref{fig:fig3}d). On the hardware used here, this nearly three-order difference in evaluation time makes repeated field predictions practical for the $512\times512$-site angle-multiplexed hologram inverse design. Even with the one-time cost of generating the training data and training DANO, this design takes 25 times less time with DANO than an estimated FDTD-based design.

Full-wave ground-truth solvers in the surveyed studies treat considerably smaller coupled arrays than the fabricated metasurfaces shown in Fig.~\ref{fig:fig5}b. Several of the device studies themselves reported that whole-device full-wave simulation was infeasible and instead simulated a reduced lens or applied the local phase approximation~\cite{she2018large,park2019allglass,li2022inverse,park2024allglass}. By reusing one learned operator across independently evaluated windows, DANO extends measured single-GPU full-field inference beyond $10^7$ meta-atoms, approaching the lower end of those fabricated-device sizes. These measurements narrow the computational gap between whole-array field evaluation and fabricated-device scale, while the hologram verified with the ground-truth solver establishes its use in inverse design at a validated scale.

\section{Discussion}\label{sec4}

Inspired by an attention-based neural operator~\cite{yu2024nonlocal}, DANO constructs its learned operator around a nonlocal physical formulation, making distributed constituent interactions an explicit learnable object. It draws complementary elements from major SciML paradigms: operator learning~\cite{li-2021-fourier-neural-operator-parametric-partial, lu2021learning, kovachki2023neural} provides reusable mappings across physical configurations, while graph and attention models~\cite{khoram2023graph} motivate explicit relational learning among constituents. Unlike PDE-residual-based approaches~\cite{zhelyeznyakov2023large}, the present one-loss training uses only a regression loss, with physics entering primarily through the architecture itself. The learned interaction allows physics-based interrogation as another validation mode, while using it to mediate full-field predictions validated from the perspective of statistical machine learning.

Conceived as an AI-for-Science tool, DANO turns the learned interaction representation into an object of scientific inquiry, rather than treating it only as an internal mechanism for prediction. Established theory on interactions can then provide representation-level validation beyond predictive accuracy. Where the effective interaction law is unknown or analytically inaccessible, the same representation could serve as a hypothesis-generating object for interaction or constitutive discovery, with resulting hypotheses subsequently examined through further analytical development or physical experiments.

DANO distills distributed interactions directly from local geometry fields without requiring a prescribed constituent parameterization or discrete interaction topology. Although the present case study uses simple rectangular constituents, the framework does not rely on this representation. This field-native construction complements graph-based approaches that explicitly prescribe entities (nodes) and connectivity (edges) for relational modeling~\cite{khoram2023graph}. In DANO, constituent representations are instead contextualized through geometry self-attention, while configuration-conditioned field-transport attention learns how field information is exchanged among them. The same formulation can extend from geometry fields of discrete constituents toward, in principle, irregular or continuous media with continuously varying material fields, although physical interpretation requires additional care.

Reusable interaction learning distilled through DANO could help narrow the growing gap between large-scale fabrication and interaction-aware modeling and design. This becomes particularly important under nonlocality, where increasing system size requires interactions among multiple constituents to be resolved jointly rather than simply replicating independent local responses. By reusing interactions learned on smaller domains, DANO enables full-field prediction and inverse design beyond the training scale, as demonstrated by the domain-size transfer in Fig.~\ref{fig:fig3} and the $512\times512$ angle-multiplexed hologram in Fig.~\ref{fig:fig5}. Fig.~\ref{fig:fig5}b further places this capability against the design--fabrication gap, where fabricated metasurfaces have reached scales beyond those routinely accessible to system-level modeling~\cite{hughes2021perspective,tsang2025fast}. Although the present implementation was not optimized for wall-clock performance, further acceleration is viable through approximate attention such as Performer~\cite{choromanski2020rethinking}, while preserving full-field prediction through domain decomposition rather than collapsing the response to task-specific outputs.

Learning nonlocal interactions through DANO requires full-field training data and therefore generally involves greater computational resources than approaches built on separation of scale, but this cost can be amortized through reuse across larger domains and repeated design queries.
We claim that the cost of training simulations should be weighed against the lifecycle cost of repeatedly resolving the coupled physics, not against a single forward solve. Full-field learning further increases the information obtained from each high-fidelity simulation: because the solver already resolves the underlying high-dimensional state field, learning directly from that field preserves substantially more information than compressing it to a few task-specific quantities of interest (QoIs), while enabling reuse across multiple observables and design objectives. We deliberately retain this field-to-field formulation even at large scale, because collapsing the response to low-dimensional QoIs would discard the spatial information through which nonlocal interactions manifest. Instead, domain decomposition addresses the scaling burden while preserving the full-field representation needed for both prediction and interaction-level physics-based interrogation. In the presented results, we address scalability by decomposing the field problem, without sacrificing the information that we feed into the model.

The present DANO formulation relies on PDE linearity to guarantee its geometry–field factorization and exact source-linear structure. The results provide strong evidence that this linear structure is indeed absorbed by the model through architectural design alone (Fig.~\ref{fig:fig5}a and Extended Data Fig.~\ref{ed:gaussian}), without additional physics-based regularization during training. 
For the systems governed by nonlinear PDEs, such as mechanical metamaterials characterized by nonlinear properties~\cite{bastek2023inverse, maurizi2025designing, zheng2026algebraic}, this factorization and source linearity generally no longer hold and the field path would require modification without overhauling. Such problems may also exhibit stronger dependence on global position, boundaries, contact, or loading history, motivating explicit conditioning on global coordinates and boundary information in addition to the local contextual field. Extending DANO beyond linear PDEs therefore would require relaxing its exact linear structure while retaining the broader principle of learning configuration-dependent distributed interactions.

DANO can also serve as an empirical probe of effective interaction range, beyond its basic role as a full-field predictor. In many nonlocal formulations, including Green's-function-based descriptions, the underlying kernel may have formally global support, yet this differs from the effective range over which distant geometry measurably changes the response. In this regard, Fig.~\ref{fig:fig3} can be interpreted as a diagnostic of context-halo sufficiency: if the dominant effective interaction range were enclosed by the prescribed halo, prediction error should approach a plateau as the surrounding domain is enlarged. Continued variation instead suggests that geometry outside the current halo still contributes to the target response. Unlike formulations that prescribe a nonlocal horizon a priori, the learned operator provides an empirical means to estimate how much context is required for a desired accuracy.

Beyond forward prediction and inverse design, DANO may help elucidate how constituent responses develop into collective nonlocal phenomena. Nonlocality is often studied through parameter sweeps in simplified, periodic settings~\cite{chen2025nonlocal}; for example, it is known that strong interactions among optical constituents can transform local resonances into collective high-$Q$ resonances~\cite{liang2024local}. Complementing mode-level and reciprocal-space analyses, DANO learns a constituent-level interaction representation from field data. Under a linear governing PDE and appropriate geometry--field factorization, this operator could be interrogated across physical parameters without prescribing how nonlocality manifests. Future work could thus move beyond the rediscovery shown in this work toward elucidating the onset of delocalized modes, roton-like dispersion, or unconventional topological states~\cite{chen2025nonlocal,wang2022nonlocal,ni2023topological,hsu2016bound}.

DANO could help close the sim-to-real gap in multiscale modeling and design~\cite{surjadi2025enabling} by representing constituent-specific fabrication uncertainty and its system-level propagation. In real-world fabrication, geometric imperfections often turn nominally periodic designs into heterogeneous realizations under aleatoric uncertainty~\cite{osanov2016topology,henneberg2020periodically,glaesener2023predicting}, especially critical for metamaterials whose functionality derives primarily from geometry rather than material composition~\cite{yu2018mechanical}. Under separation of scale, computational robust design has often treated constituent perturbations with periodic boundary conditions, using a prescribed~\cite{wang2019robust} or learned~\cite{chen2023gan} uncertainty representation. Under nonlocality, imperfections affect distributed interactions and collective response across constituents. By retaining constituent-specific geometry and physical context, DANO could capture realization-specific imperfections beyond nominal parameterizations, supporting system-level robust design. Our case study provides a simplified analog: nominally square-shaped constituents vary in their two side lengths across the array, while vacancies represent missing-constituent defects.

A remaining limitation is the computational overhead associated with very large effective interaction ranges. If the context required for accurate prediction grows with the physical domain, the number of constituent tokens can become large, particularly for full-field prediction in aperiodic systems. The present study prioritizes the network architecture design principle rather than runtime optimization, but several directions could reduce this burden, including approximate attention, adaptive or error-controlled context truncation, multi-resolution tokenization, and hierarchical operator evaluation. These strategies would become especially useful when interactions decay slowly or when the effective interaction range approaches the full system size.

\section{Methods}\label{sec3}

We describe the physical system and full-wave data generation in the presented case study, the DANO formulation, the training and evaluation procedures, and the inverse-design and full-wave verification protocols.

\subsection{Physical system and dataset generation}

\bmhead{Physical system and learning target}
We considered electromagnetic scattering from finite, aperiodic arrays of rectangular silicon nitride (SiN) pillars patterned on a silica substrate. Each array comprised $33\times33$ lattice sites with a pitch of 300\,nm, spanning $9.9\times9.9$\,\textmu m$^2$. The pillars were 600\,nm tall. Their cross-sections were drawn from a library of $41\times41$ side-length pairs $(L_x,L_y)$ from 50 to 250\,nm at 5-nm increments, and each pillar was sampled independently and uniformly from this library. Each site was occupied independently with probability 0.8, so that both pillar-size disorder and vacancies contributed to the geometric heterogeneity. The pillars stood on a silica substrate with refractive index $n_{\mathrm{sub}}=1.45$, with air above and between them. The pillar refractive index was $n=2.106$ at 532\,nm, with no absorption. The simulated stack and the conversion from full-wave simulations to local training windows are summarized in Extended Data Fig.~\ref{ed:data}a,d.

All calculations were performed at the free-space wavelength $\lambda=532$\,nm using the $\exp(-\mathrm{i}\omega t)$ convention. Positions are $\mathbf{r}=(x,y,z)$, with $x$ and $y$ along the lattice axes in the array plane and $z$ normal to the substrate, pointing into air. The electric field obeys
\begin{equation}
\nabla\times\nabla\times \mathbf{E}(\mathbf{r})-k_0^2\varepsilon(\mathbf{r})\mathbf{E}(\mathbf{r})=0,
\qquad k_0=2\pi/\lambda,
\label{eq:m-wave}
\end{equation}
where $\varepsilon=\varepsilon_{\mathrm{bg}}+\Delta\varepsilon_g$ separates the planar air-substrate background from the contrast introduced by a layout $g$. For each source, we decomposed the total field as $\mathbf{E}=\mathbf{E}_b+\mathbf{E}_s$, where $\mathbf{E}_b$ is the field produced by the same source in the bare layered stack and $\mathbf{E}_s$ is the constituent-induced scattered field. At fixed layout, scattering is complex-linear in the background field, and we write
\begin{equation}
\mathbf{E}_s=\mathcal{T}[g]\mathbf{E}_b.
\label{eq:m-operator}
\end{equation}
The operator $\mathcal{T}[g]$ depends nonlinearly on the layout but linearly on the illumination. This geometry--field separation defines the learning task used throughout the study.

\bmhead{Full-wave ground-truth simulations}

Ground-truth fields were generated by three-dimensional finite-difference time-domain (FDTD) simulations in Ansys Lumerical FDTD 2024 R2, using a 20-nm mesh in the array plane and perfectly matched layer (PML) boundaries. Plane-wave illumination from the substrate was implemented with a total-field/scattered-field (TFSF) source to suppress finite-source edge diffraction at oblique incidence. The simulation geometry is shown in Extended Data Fig.~\ref{ed:data}a. On an observation plane 350\,nm above the pillar tops, we recorded the total field $\mathbf{E}$ and the bare-stack background field $\mathbf{E}_b$ on the same $496\times496$ grid. From these we extracted the in-plane components $E_{s,x}$ and $E_{s,y}$ of the scattered field $\mathbf{E}_s=\mathbf{E}-\mathbf{E}_b$, whose real and imaginary parts formed the four learning targets. The normal component $E_{s,z}$ was not extracted because, in the homogeneous air above the pillars, it follows from $E_{s,x}$ and $E_{s,y}$ through $\nabla\cdot\mathbf{E}_s=0$. Source-only FDTD fields supplied $\mathbf{E}_b$. For arrays without such a simulation, $\mathbf{E}_b$ was written in closed form.

\bmhead{Illumination corpora and data partitions}
We generated 426 plane-wave cases across 33 substrate-side incidence conditions: normal incidence and polar angles of $10^\circ$, $20^\circ$, $30^\circ$ and $40^\circ$, each at eight azimuths. The in-plane incident polarization was aligned with the sample $x$ axis. Each case pairs one layout with one illumination. A second corpus comprised 3,600 structured-illumination cases, obtained by pairing each of 300 band-limited speckle realizations with 12 independently sampled layouts. The speckle fields were synthesized from random $s$- and $p$-polarized angular-spectrum components within the substrate propagation cone (Extended Data Fig.~\ref{ed:data}b,c).

The plane-wave cases were divided into 355 training and 71 validation cases, and the speckle cases into 3,000 training and 600 validation cases. Layouts were disjoint between training and validation, whereas the incidence conditions and speckle realizations were represented in both. Validation data guided model selection. In-distribution accuracy was measured on the separate test set described under ‘Evaluation protocol’. The test and transfer cases were not used for model development.

\subsection{DANO formulation and architecture}

\bmhead{Windowed scattering operator}
In multiple-scattering form, the configuration-dependent response can be written as
\begin{equation}
\mathcal{T}[g]=\mathbb{G}_{\mathrm{out}}\mathbb{T}\left(\mathbb{I}-\tilde{\mathbb{G}}\mathbb{T}\right)^{-1},
\label{eq:m-master}
\end{equation}
where $\mathbb{T}$ collects constituent transition operators, $\tilde{\mathbb{G}}$ propagates responses between distinct sites through the layered background and $\mathbb{G}_{\mathrm{out}}$ maps constituent responses to the observation plane. This factorization motivates, but is not imposed exactly on, the learned architecture.

DANO approximates $\mathcal{T}[g]$ through local window operators that share one set of learned parameters. Let $W_k$ denote the subregion of the array covered by a prediction core and its context halo, and let $P_k$ apply the core's partition-of-unity weight and extend its prediction by zero to the full array. The assembled operator is
\begin{equation}
\mathcal{T}_{\mathbf{w}}[g]\mathbf{E}_b
=\sum_k P_k\left[
\mathcal{T}_{\mathbf{w}}[g|_{W_k}]\,
\mathbf{E}_b|_{W_k}
\right],
\label{eq:m-hypothesis}
\end{equation}
where $\mathbf{w}$ denotes the learned parameters. The finite input window bounds the spatial context used to predict the field on each core, while the window operator models nonlocal interaction among the constituents within that context.

\bmhead{Windows and context halo}
Each window contained a prediction core of about $4\times4$ constituent spacings ($62\times62$ grid points, $1.24\times1.24$\,\textmu m$^2$) surrounded by a context halo of 7 spacings (2.1\,\textmu m) on each side. The geometry encoder received a permittivity crop of $288\times288$ grid points comprising the prediction core, the context halo and an additional 160-nm margin on each side. Extended Data Fig.~\ref{ed:window}a,b illustrates the prediction core, context halo, geometry crop and independent window predictions. Each occupied or vacant lattice site in the window was represented by a token at its site center, with both types processed by the same geometry and drive encoders. The halo width was selected by retraining models with widths from 1 to about 13 spacings (0.3 to 4\,\textmu m) under otherwise fixed settings (Extended Data Fig.~\ref{ed:window}c).

\bmhead{Geometry representation}

The geometry input was a permittivity map on the 20-nm grid in which each cell holds its SiN area fraction: zero in air, one inside a pillar and fractional in cells cut by a pillar boundary. Three discrete–continuous (DISCO) convolutions \cite{liuschiaffini-2024-neural-operators-with-localized-integral} mapped each crop to a spatial feature map. Their kernels were parameterized continuously in physical coordinates on a 60-nm-radius support, allowing the learned coefficients to be resampled at different mesh spacings (Extended Data Fig.~\ref{ed:invariants}a). The feature map was sampled bilinearly at lattice-site centers and projected to 128 dimensions.

The geometry path comprised eight self-attention layers of width 128, each with four heads and a residual multilayer perceptron \cite{vaswani-2017-attention-is-all-you-need}. Two-dimensional rotary embeddings with learned frequencies encoded token positions \cite{su-2024-roformer-enhanced-transformer-with-rotary}. The attention and multilayer-perceptron residual branches used zero-initialized ReZero gains \cite{bachlechner-2020-rezero-is-all-you-need}. The geometry path received no illumination field. Linear heads on its per-token embeddings generated the transport attention patterns, complex local gains and decoder projections used by the field path.

\bmhead{Linear field path}
The field path (Fig.~\ref{fig:fig2}b) applies the configuration-conditioned operator to the drive. At each token, the drive comprised the background field and its first- and second-order transverse derivatives,

\begin{equation}
\left[
\mathbf{E}_b,\ 
k_0^{-1}\partial_x\mathbf{E}_b,\ 
k_0^{-1}\partial_y\mathbf{E}_b,\ 
k_0^{-2}\partial_{xy}\mathbf{E}_b,\ 
k_0^{-2}\partial_{xx}\mathbf{E}_b,\ 
k_0^{-2}\partial_{yy}\mathbf{E}_b
\right].
\label{eq:m-drive}
\end{equation}

Applied to all three Cartesian field components, this yielded 18 complex channels per token. The derivatives were approximated by central finite differences on the stored 20-nm grid. These channels are fixed linear functions of $\mathbf{E}_b$ and provide information about field variation between the 300-nm-spaced tokens, mitigating aliasing from token-level sampling \cite{papoulis-1977-generalized-sampling-expansion}. A bias-free complex-linear layer lifted the drive to a 256-dimensional complex state.

The field path contained eight interaction layers with no activation functions, normalization layers or biases. Let $b$ denote the stacked lifted drive and $h^{(\ell)}$ the token states, with $h^{(0)}=A_0b$. Each layer computes

\begin{equation}
h^{(\ell)}=S_\ell\left(b+\beta_\ell G_b^{(\ell)}h^{(\ell-1)}\right),
\qquad
S_\ell=(I+\gamma_\ell M_\ell)A_\ell,
\label{eq:m-layer}
\end{equation} where $A_\ell$ multiplies the 256 channels of token $i$ elementwise by the configuration-conditioned complex gains $\alpha[g_i^{(\ell)}]\in\mathbb{C}^{256}$ computed from its geometry embedding $g_i^{(\ell)}$ and $M_\ell$ is a $256\times256$ complex matrix applied at every token. The learned scalar gates $\beta_\ell$ and $\gamma_\ell$ are initialized to zero. The transport operator has blocks

\begin{equation}
G_{b,ij}^{(\ell)}
=D(r_i)
\left(
\sum_{h=1}^{4}t_{ij}^{(\ell,h)}\Pi_h
\right)
D(\overline{r_j})V_\ell,
\label{eq:m-transport}
\end{equation} where $t_{ij}^{(\ell,h)}$ are softmax transport patterns determined by the geometry embeddings and token positions, $\Pi_h$ selects the channel slice of head $h$, and $V_\ell$ is a complex channel map. The conjugate factors $D(r_i)$ and $D(\overline{r_j})$, with $r_i=e^{i\varphi_i}$, $\varphi_i=k_0\boldsymbol{\rho}_i\cdot\boldsymbol{f}$, learned frequencies $\boldsymbol{f}$ and window-relative token position $\boldsymbol{\rho}_i$, introduce the pairwise phase $e^{i(\varphi_i-\varphi_j)}$ (phase rotor in Fig.~\ref{fig:fig2}b). $S_\ell$ and $G_b^{(\ell)}$ are the learned counterparts of $\mathbb{T}$ and $\tilde{\mathbb{G}}$ in Eq.~\eqref{eq:m-master}, the local-response and transport blocks of Fig.~\ref{fig:fig2}b, and the eight-layer recurrence unrolls the inverse in Eq.~\eqref{eq:m-master}.

All coefficients in these updates depend only on geometry and position, so every state $h^{(\ell)}$ is complex-linear in $\mathbf{E}_b$. Spectral normalization \cite{miyato-2018-spectral-normalization-for-generative-adversarial} controls the gain of the complex-linear weights without introducing illumination-dependent normalization.

\bmhead{Decoder and full-field assembly}
Two configuration-conditioned cross-attention layers mapped the final token states to arbitrary query positions inside the prediction core. Query embeddings were constructed from window-relative positions. The score of a query--token pair was the raw inner product of their rotary-encoded projections, without softmax normalization, so the decoder formed an additive sum of token contributions. A final bias-free complex-linear head emitted $(E_{s,x},E_{s,y})$ at the query points. DANO contained 3.86 million real scalar parameters, counting the real and imaginary components of complex weights separately.

At inference, each core was evaluated at every valid grid point. Core origins were placed on an overlapping grid with at least 320\,nm of overlap. Window outputs were combined using a $\cos^2$ partition-of-unity weight normalized to sum to one wherever cores overlapped. Because the assembly is a linear weighted sum, the full-field map remains complex-linear in the illumination and supports gradient-based optimization of pillar dimensions. Extended Data Fig.~\ref{ed:window}d--f shows the assembled error, overlap weights and one seam-level comparison with FDTD.

\bmhead{Mesh transfer and linearity tests}
The DISCO kernels and rotary phases were defined in physical coordinates. For evaluation on finer transverse meshes, we kept the learned parameters and physical window dimensions fixed and sampled the continuous geometry kernels at the new grid spacing. The model was not fine-tuned using fine-mesh data. Extended Data Fig.~\ref{ed:invariants}b--d reports transfer from the 20-nm training mesh to 12.5- and 9.375-nm meshes.

At a fixed layout, the architecture is complex-linear in $\mathbf{E}_b$: the prediction for a complex linear combination of illuminations equals the same combination of their predictions, up to numerical precision. A superposition check in single precision gave a relative deviation of $7.9\times10^{-7}$.

\subsection{Model training and evaluation}

\bmhead{Training samples and objective}

Training used local windows rather than complete arrays. For each draw, a case was selected from the structured-illumination corpus with probability 0.48 and from the plane-wave corpus otherwise. We randomly shifted the core grid by up to half a core width in each in-plane direction, so a given spatial location appeared at different positions within the prediction core across draws. The shifts also exposed the model to different local fields within each speckle realization (Extended Data Fig.~\ref{ed:data}d--f). For each window, we sampled 1,024 valid query points. Batches contained eight windows, with samples drawn online rather than organized into epochs.

For window $w$, let $Q_w$ denote the sampled valid query points. The training objective was the mean relative squared error

\begin{equation}
\mathcal{L}
=\frac{1}{B}\sum_{w=1}^{B}
\frac{
\sum_{\mathbf{q}\in Q_w}\lVert\widehat{\mathbf{E}}_s(\mathbf{q})-\mathbf{E}_s(\mathbf{q})\rVert_2^2
}{
\max\left(
\sum_{\mathbf{q}\in Q_w}\lVert\mathbf{E}_s(\mathbf{q})\rVert_2^2,\ 
\epsilon\,|Q_w|
\right)
},
\label{eq:m-loss}
\end{equation} with $B=8$ and $\epsilon=1.31\times10^{-7}$. The energy floor prevented windows with very weak scattered fields from receiving arbitrarily large relative weight. Each window contributed equally to the batch loss. Input fields were used at their stored amplitudes without per-window normalization.

\bmhead{Training protocol}
The model was optimized for 240,000 steps using AdamW \cite{loshchilov-2019-decoupled-weight-decay-regularization}, with an initial learning rate of $6\times10^{-4}$, weight decay $10^{-5}$ and gradient-norm clipping at 1.0. The learning rate was increased linearly over the first 500 steps and followed a cosine decay to 5\% of its initial value over the final 60,000 steps. An exponential moving average of the weights with decay 0.999 was maintained during training.

Validation was performed every 5,000 steps on 16 fixed cases per corpus. For each of five training seeds, the best exponential-moving-average checkpoint on these cases was evaluated once on the test set to assess training-seed variability. All other analyses and the inverse design used one of these checkpoints.

\bmhead{Evaluation protocol}

Full-field accuracy was quantified using normalized mean squared error (NMSE),

\begin{equation}
\operatorname{NMSE}
=
\frac{
\sum_{\mathbf{x}\in\Omega}
\lVert\widehat{\mathbf{E}}_s(\mathbf{x})-\mathbf{E}_s(\mathbf{x})\rVert_2^2
}{
\sum_{\mathbf{x}\in\Omega}
\lVert\mathbf{E}_s(\mathbf{x})\rVert_2^2
},
\label{eq:m-nmse}
\end{equation} where $\Omega$ is the set of grid points on the observation plane outside a 540-nm perimeter strip, and the norm includes both transverse field components. NMSE was computed for each case, and case-level values were averaged for accuracy comparisons.

Field errors were computed from a non-overlapping $8\times8$ grid of prediction cores on the $496\times496$ plane. The overlapping partition-of-unity assembly was used for displayed fields, inverse design and the uniform-array, mesh-transfer and larger-array evaluations.

In-distribution accuracy was evaluated on a separate test set. It contained 213 plane-wave and 600 structured-illumination cases with newly sampled layouts under the incidence conditions and speckle realizations represented in training. Validation data guided checkpoint selection, halo-width selection and baseline learning-rate selection. Structured-layout transfer was evaluated on a hyperbolic metalens with a 10-\textmu m focal length at three incidence angles. Illumination transfer comprised 18 cases at held-out incidence settings, 18 coherent plane-wave superpositions and four Gaussian beams. Mesh transfer used independent FDTD solutions for 50 layouts at each of the two finer meshes. Domain-size transfer used the same checkpoint and physical window dimensions for one layout at each array size from $64^2$ to $768^2$ sites.

\bmhead{Source-band decomposition}

At a fixed layout, the scattered field is linear in the incident field, and DANO is complex-linear in its drive, so the fields and predictions of a speckle source split into source bands sum over the bands. Each source band collected the angular components of the speckle source at the injection plane within a range of $f=|\mathbf{k}_\parallel|/k_0$, where $\mathbf{k}_\parallel$ is the in-plane wavevector, without renormalization. Angular components with $f>1$ propagate in the silica substrate but not in air. Source bands were simulated for the bare stack and the array with the solver settings of the corpora. The two-band split at $f=1$ used 32 randomly selected test realizations, each on its two test layouts, and instead obtained the $f\le1$ band by subtraction from the full-spectrum fields. The eight-band profile used 12 realizations, each on the first of its two test layouts. The share of $f>1$ is the case-averaged $f>1$ term of the NMSE plus half of the case-averaged cross term, divided by the case-averaged NMSE.

\bmhead{Benchmarking}

We compared DANO with a U-Net \cite{ronneberger2015u}, a Fourier neural operator (FNO) \cite{li-2021-fourier-neural-operator-parametric-partial}, a multiple-input operator network (MIONet) \cite{jin-2022-mionet-learning-multiple-input-operators-tensor} and a neural Green's operator (NGO) \cite{melchers-2026-neural-greens-operators-parametric-partial} (Fig.~\ref{fig:fig3}e and Extended Data Fig.~\ref{ed:gaussian}). The baselines used the same window draws, corpus splits, query regions and 240,000 training steps as DANO. They contained 3.76–4.20 million real scalar parameters, compared with 3.86 million for DANO. Each baseline received eight input channels—the permittivity, a domain mask marking the simulated region within the window, and the real and imaginary parts of the three Cartesian components of $\mathbf{E}_b$—and predicted the two transverse components of $\mathbf{E}_s$ on the same core.

The baselines differ in how they combine the permittivity with the background field. U-Net and FNO process all input channels together on the window, through local convolutions and global Fourier layers, respectively. MIONet encodes the permittivity and the background field in separate branches and combines them with a position trunk. NGO followed the data-driven form of its original study \cite{melchers-2026-neural-greens-operators-parametric-partial}, in which a convolutional network maps the permittivity to a complex operator that acts on the background field. NGO is therefore complex-linear in $\mathbf{E}_b$, as DANO is, but represents the permittivity through a fixed global basis rather than through interactions among discrete constituents.

To account for the large difference in background-field amplitudes between the two illumination corpora, the six background-field channels of U-Net, FNO and MIONet were divided by the window’s root-mean-square background amplitude, and the predicted scattered field was rescaled by the same factor. The permittivity and domain-mask channels were left unscaled. This wrapper contained no learned parameters. Applying the corresponding drive and output rescaling to DANO or NGO leaves the prediction unchanged because both models are complex-linear in $\mathbf{E}_b$, so NGO was trained without it. Each baseline used one learning rate for both corpora. It was selected from $6\times10^{-5}$, $2\times10^{-4}$, $6\times10^{-4}$ and $2\times10^{-3}$, with $2\times10^{-5}$ added for U-Net, by the lowest validation-subset error of the model trained on both corpora. The selected baselines trained on both corpora were also run with the five training seeds of DANO, with the data split fixed.
All models were evaluated using the same full-field NMSE on the same test, validation and transfer cases. Extended Data Fig.~\ref{ed:gaussian} reports their Gaussian-beam transfer.

\subsection{Physics-based interrogation of the learned operator}
\label{sec:m-operator-analysis}

Following the two response directions in Eq.~\eqref{eq:two_kernels}, we examined changes in the learned scattered field when varying the drive at fixed layout and when varying individual pillars at fixed drive. The analyses below use the trained window operator $\mathcal T_{\mathbf w}[g]$ and controlled geometry perturbations, respectively.

\bmhead{Localized drive probes and radiation reference}

To examine the drive-response direction introduced in Eq.~\eqref{eq:two_kernels}, we held the layout fixed and applied $x$- and $y$-polarized Gaussian packets $\psi_j(\mathbf r)=\exp[-|\mathbf r-\mathbf r_j|^2/(2\sigma^2)]$, centered at constituent sites with $\sigma=300$\,nm. Packet values and their first- and second-order derivatives supplied consistent drive features. The packets were used as diagnostic inputs without enforcing Maxwell transversality.

We compared the learned responses with a sum of independent pillar responses above an air–glass interface, weighted by the same packet envelope and evaluated on the observation plane. Each pillar response was taken from full-wave simulations of an isolated pillar of the same size and represented by electric-dipole, magnetic-dipole and electric-quadrupole sources. Radial profiles were computed by annular averaging, and angular contrast was defined as the response magnitude divided by its azimuthal mean at each radius.

\bmhead{Phase-ramp commutation test}

For one fixed window and its background field $\mathbf{E}_b$, we compared $\mathcal T_{\mathbf{w}}[g](M_k\mathbf{E}_b)$ with $M_k\,\mathcal T_{\mathbf{w}}[g](\mathbf{E}_b)$, corresponding to applying a phase ramp before or after the learned operator (Eq. \eqref{eq:commutation_metric}). Derivative input features were updated consistently with the ramp. The commutation residual was the norm of the difference divided by the norm of the unmodulated output.

\bmhead{Configuration-response measurements and spatial range}

To probe the configuration-response direction represented by $J[a;f]$ in Eq.~\eqref{eq:two_kernels}, we held the drive $f$ and background $a_0$ fixed while changing one pillar at a time. The background field $\mathbf{E}_b$ therefore remained unchanged, so changes in the total and scattered fields were identical. We analyzed three fully occupied $19\times19$ layouts: disordered, uniform and a hyperbolic metalens. For pillar $j$, let $g_j^\pm$ denote layouts in which both side lengths were increased or decreased by 20\,nm. Using identical window and query positions, we measured
\begin{equation}
  \Delta_j\widehat{\mathbf E}_s
  =\frac{1}{2}
  \left[
    \mathcal T_{\mathbf w}[g_j^+]\,\mathbf{E}_b
    -\mathcal T_{\mathbf w}[g_j^-]\,\mathbf{E}_b
  \right].
  \label{eq:m-configuration-response}
\end{equation}

These finite pillar-size changes give a finite-difference counterpart of the functional derivative $J[a;f]$. Radial response profiles were averaged over displacement annuli and normalized by the response at the perturbed site.

To quantify how much of the measured response lay beyond a radius $R$, let $\Delta E_{ij}$ denote a sampled component of the response at query point $i$ to a change at pillar $j$, and let $r_{ij}$ be the in-plane distance between query point $i$ and pillar $j$. We defined
\begin{equation}
  \epsilon(R)=
  \left[
    \frac{\sum_{r_{ij}>R}|\Delta E_{ij}|^2}
         {\sum_{ij}|\Delta E_{ij}|^2}
  \right]^{1/2},
  \qquad
  R_\tau=\inf\{R:\epsilon(R)\leq\tau\}.
  \label{eq:m-finite-response-tail}
\end{equation}
$R$ was evaluated up to 3\,\textmu m. For the drive curve in Fig.~\ref{fig:fig4}f, $\epsilon(R)=\lVert\mathcal T_{\mathbf w}[g]\mathbf{E}_b-\mathcal T^{(R)}_{\mathbf w}[g]\mathbf{E}_b\rVert/\lVert\mathcal T_{\mathbf w}[g]\mathbf{E}_b\rVert$, where $\mathcal T^{(R)}_{\mathbf w}[g]$ omits token contributions farther than $R$ from each query point.

\subsection{Inverse design and full-wave verification}

\bmhead{Design task and parameterization}

We used the trained DANO model, with its parameters fixed, to design a fully occupied $512\times512$-site metasurface for angle-multiplexed holography. The substrate-side input was an $x$-polarized plane wave at 532\,nm, incident at either $0^\circ$ or $20^\circ$ in the $x$–$z$ plane. The respective targets were the left half of the POSTECH seal and the right half of the Northwestern logo, occupying disjoint pixels of the far-field target region.

Far-field intensity was computed from the transverse total field on the observation plane,
\begin{equation}
I(\mathbf{k}_{\parallel})=
|\mathcal{F}E_x|^2+|\mathcal{F}E_y|^2,
\label{eq:m-farfield}
\end{equation}
and scored within the propagating cone $|\mathbf{k}_{\parallel}|\leq k_0$. The target region was translated with the incident in-plane wavevector so that both angles used the same coordinates relative to the incident direction.

Each pillar side length $L_{\alpha,s}$, for $\alpha\in\{x,y\}$, was parameterized as $L_{\alpha,s}=[50+200\sigma(u_{\alpha,s})]$\,nm, where $u_{\alpha,s}$ was unconstrained and $\sigma$ was the logistic function. Side lengths were initialized independently and uniformly between 120 and 180\,nm. We constructed the permittivity map on the 20-nm grid from the pillar dimensions with the same rule as the training input, in which each grid cell holds its SiN area fraction, so the map varies continuously with the side lengths. The background field and the assembled DANO prediction gave the total field $\mathbf{E}_b+\widehat{\mathbf{E}}_s$, from which we computed the far-field objective and differentiated it with respect to the pillar dimensions.

\bmhead{The inverse design problem}

The DANO and local designs solve the same optimization problem with different forward models $m\in\{\mathrm{DANO},\mathrm{local}\}$,
\begin{equation}
\begin{aligned}
\min_{\mathbf{p}}\quad
& \frac{1}{2}\sum_{\theta\in\{0^\circ,20^\circ\}}
\mathcal{L}^{(m)}_\theta(\mathbf{p}) \\
\text{s.t.}\quad
& 50\,\mathrm{nm}\le L_{\alpha,s}\le 250\,\mathrm{nm}.
\end{aligned}
\label{eq:m-design-problem}
\end{equation}
where $\mathbf{p}=\{L_{\alpha,s}\}$ collects the two side lengths of all $512^2$ pillars. DANO gives the observation-plane field $\mathbf{E}_\theta=\mathbf{E}_{b,\theta}+\mathcal{T}_{\mathbf{w}}[g(\mathbf{p})]\mathbf{E}_{b,\theta}$ with its trained parameters $\mathbf{w}$ fixed. The local model assigns each site the complex transmission of its periodic unit cell at angle $\theta$, taken from a library of 1{,}681 cells. In both cases, $I_\theta(\mathbf{p})$ follows from Eq.~\eqref{eq:m-farfield}. With the propagating cone $C=\{|\mathbf{k}_\parallel|\le k_0\}$ and the target region $\Omega_\theta\subset C$, which moves with the incident in-plane wavevector,
\begin{equation}
\begin{aligned}
\mathcal{L}^{(m)}_{\theta}={}&
1-\operatorname{NCC}_{C}(I_{\theta},Y_{\theta})
+\lambda_{\eta}\max(0,\eta_{\min}-\eta_{\theta})^2\\
&+\lambda_{\times}\max\!\left(0,\operatorname{NCC}_{\Omega_\theta}(I_{\theta},Y_{\theta'})\right)
+\lambda_{\Pi}\max(0,\Pi_{\min}-\Pi_{\theta})^2,
\end{aligned}
\label{eq:m-design}
\end{equation}
where $\theta'$ is the other angle and $\operatorname{NCC}_A$ is the zero-mean normalized cross-correlation over $A$. Each target $Y_\theta$ is zero outside $\Omega_\theta$, the target-power fraction is $\eta_\theta=\sum_{\Omega_\theta}I_\theta/\sum_{C}I_\theta$, and $\Pi_\theta$ is the mean of $|E_{\theta,x}|^2+|E_{\theta,y}|^2$ over the array relative to that of the background field. Both designs used $\eta_{\min}=0.02$, $\lambda_{\eta}=1{,}000$ and $\lambda_{\times}=1$. DANO also used $\Pi_{\min}=0.55$ and $\lambda_{\Pi}=10$. The local model, which does not resolve the array field, omitted the $\Pi_\theta$ term and computed image matching over $\Omega_\theta$.

\bmhead{Optimization implementation}

For DANO, the bounds were imposed by the logistic parameterization above, and we minimized over $u_{\alpha,s}$ for 600 iterations using Adam \cite{kingma-2015-adam-a-method-for-stochastic} with a learning rate of 0.02. The DANO design in Fig.~\ref{fig:fig5}a came from a single run without selection among candidates. The local design was optimized with Adam directly on the side lengths for 2{,}000 iterations.

\bmhead{Full-wave ground-truth verification} The optimized DANO and local-baseline layouts were exported as pillar lists and independently simulated with the same finite-array FDTD setup at both incidence angles. Their far-field images were evaluated using identical Fourier conventions, propagating-cone masks, target regions and targets. For each angle, we computed the normalized cross-correlation $\operatorname{NCC}_{\Omega_\theta}$ of the target-region image with its intended target, its correlation with the other target, and the difference between the two. We also computed the correlation between the two target-region images in coordinates translated with the incident wavevector. DANO-predicted scores are given for comparison, and design verification used the FDTD scores.
\section*{Data Availability}

The full-wave field data supporting the findings of this study are available from the corresponding author upon reasonable request.

\section*{Code Availability}

The source code for DANO, including data preprocessing, model training, inference, and inverse-design demonstrations, is available from the corresponding author upon reasonable request.

\section*{CRediT author statement}
Dex Doksoo Lee: Conceptualization, Methodology, Writing -- Original Draft, Writing -- Review \& Editing, Visualization, Project Administration. 
Jehyeon Shin: Conceptualization, Methodology, Software, Validation, Formal Analysis, Investigation, Data Curation, Writing -- Original Draft, Writing -- Review \& Editing, Visualization. 
Yue Yu: Methodology, Writing -- Review \& Editing. 
Sunkyu Yu: Methodology, Writing -- Review \& Editing. 
Junsuk Rho: Writing -- Review \& Editing, Supervision, Project Administration, Funding Acquisition. 
Wei Chen: Writing -- Review \& Editing, Supervision, Project Administration, Funding Acquisition. 

\bmhead{Acknowledgements}
W.C. and D.D.L. acknowledge support from the U.S. National Science Foundation (NSF) Boosting Research Ideas for Transformative and Equitable Advances in Engineering (BRITE) Fellow Program (CMMI-2227641) and NASA's Minority University Research and Education Project Institutional Research Opportunity (MIRO) through the Center for In-Space Manufacturing (CISM-R2): Recycling and Regolith Processing (Award 80NSSC24M0176). W.C. also acknowledges support from the NSF Materials Research Science and Engineering Centers (MRSEC) program (DMR-2308691). Y.Y. acknowledges support from the National Institutes of Health (NIH) under award 1R01GM157589-01 and the NSF under award 2427915. J.R. acknowledges the POSCO-POSTECH-RIST Convergence Research Center program funded by POSCO, and the National Research Foundation (NRF) grant (RS-2024-00356928) funded by the Ministry of Science and ICT (MSIT) of the Korean government. J.S. acknowledges the Hyundai Motor Chung Mong-Koo Foundation; the Institute of Information \& Communications Technology Planning \& Evaluation (IITP) (grant no. RS-2019-II191906; POSTECH Artificial Intelligence Graduate School Program) funded by the Ministry of Science and ICT (MSIT) of the Korean government; and the National Research Foundation of Korea (NRF) Ph.D. fellowship (grant no. RS-2025-25438817) funded by the Ministry of Education (MOE) of the Korean government.

\section*{Competing Interests}
The authors declare no competing financial or non-financial interests.
\bibliography{sn-bibliography}
\clearpage
\setcounter{figure}{0}
\renewcommand{\theHfigure}{ED\arabic{figure}}
\renewcommand{\figurename}{Extended Data Fig.}

\begin{figure}[h!]
    \centering
    \includegraphics[width=1.0\linewidth]{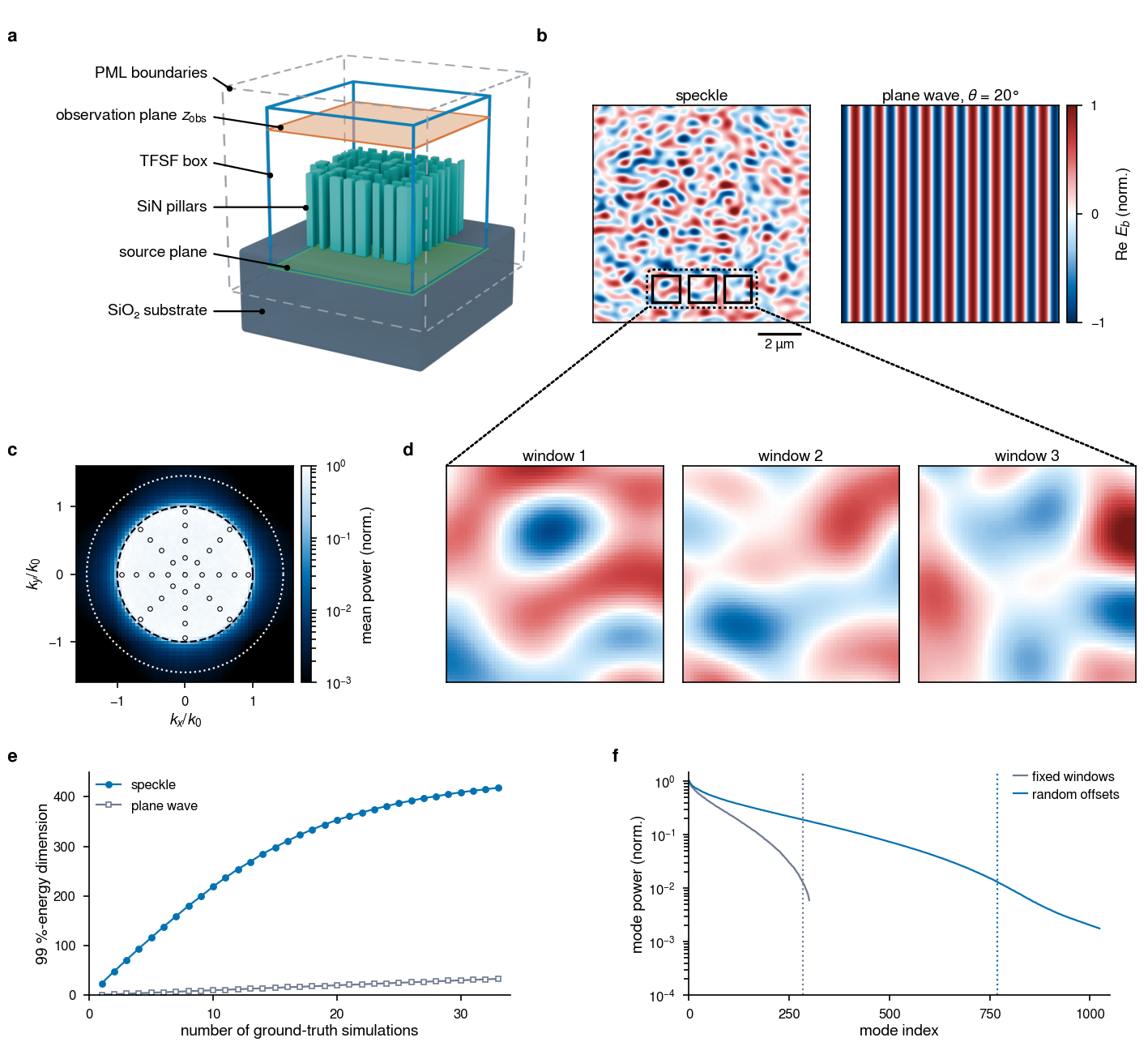}
    \caption{\textbf{Full-wave data generation and window-level illumination diversity.} \textbf{a}, Simulation stack with a total-field/scattered-field (TFSF) source plane and box, observation plane $z_{\mathrm{obs}}$ and perfectly matched layer (PML) boundaries. Vertical distances are not to scale. \textbf{b}, Real part of the background field $\mathbf{E}_b$ for one speckle realization and a plane wave at $\theta=20^\circ$, each normalized by its 99.5th-percentile magnitude. Solid squares mark the three $1.24$-\textmu m windows of \textbf{d} (scale bar, 2\,\textmu m). \textbf{c}, Background-field power spectrum averaged over 300 speckle realizations (logarithmic color scale). Dashed and dotted circles, air and substrate light cones; open circles, the 33 plane-wave settings. \textbf{d}, Background fields in the three windows of \textbf{b}. \textbf{e}, 99\%-energy dimension, the number of modes holding 99\% of the background-field energy in local windows, versus the number of ground-truth simulations. \textbf{f}, Normalized Karhunen–Loève mode spectra of speckle background fields in fixed windows and in windows with random offsets. Dotted lines, the 99\%-energy dimensions.}
    \label{ed:data}
\end{figure}
\clearpage

\begin{figure}[h!]
    \centering
    \includegraphics[width=1.0\linewidth]{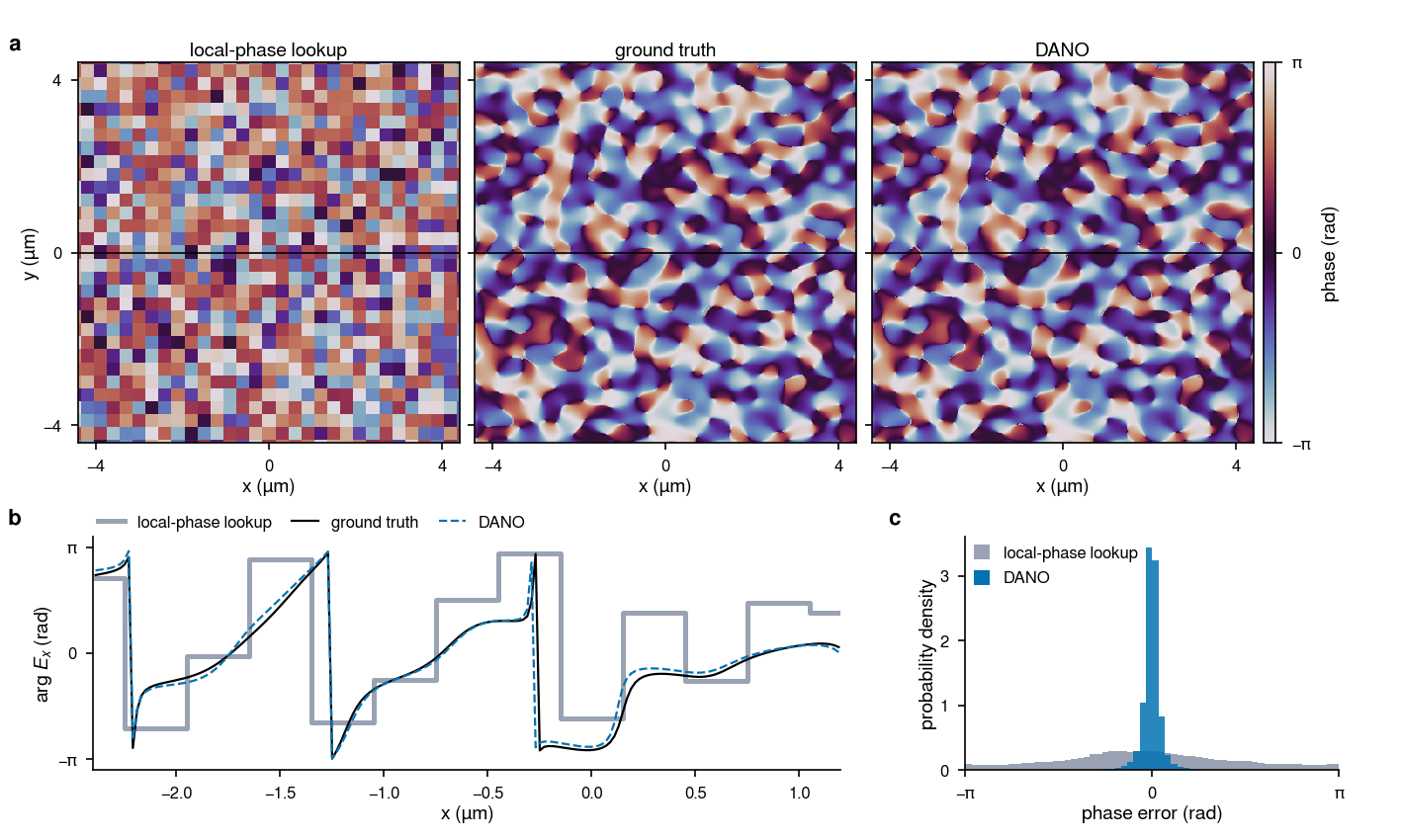}
    \caption{\textbf{Phase mismatch between the local phase approximation and the ground truth.} One held-out, fully occupied random layout under normal-incidence illumination, evaluated on the observation plane. \textbf{a}, Phase of the transmitted field $E_x$ from the local phase approximation (LPA, computed by a local-phase lookup, left), the ground truth from finite-difference time-domain simulation (middle) and domain-decomposition attention neural operator (DANO, right). The black line marks the row examined in \textbf{b}. \textbf{b}, Phase along the marked row. \textbf{c}, Distributions of circular phase error relative to the ground truth.}
    \label{ed:lpa}
\end{figure}
\clearpage

\begin{figure}[h!]
    \centering
    \includegraphics[width=1.0\linewidth]{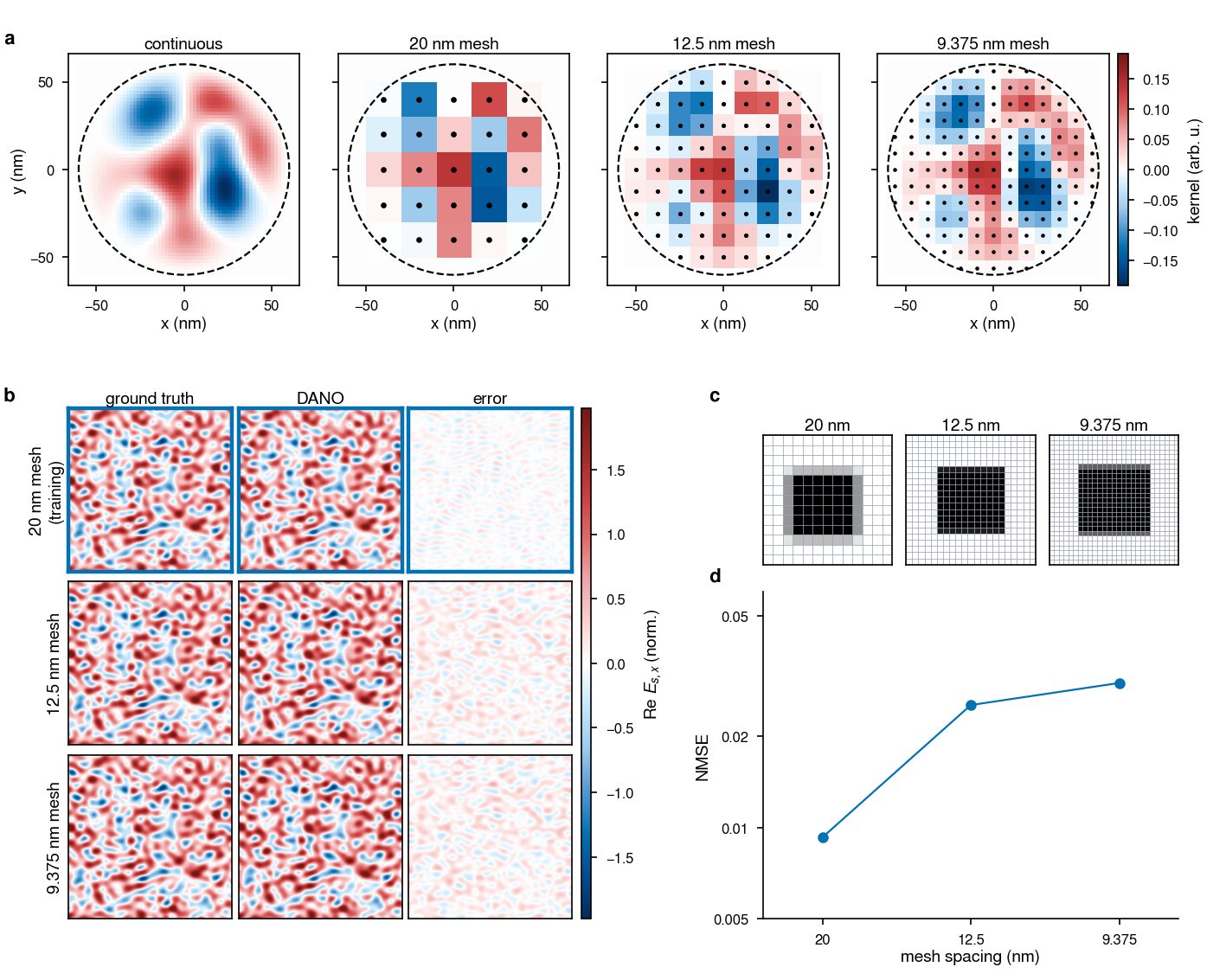}
    \caption{\textbf{Full-field prediction on finer simulation meshes.} The domain-decomposition attention neural operator (DANO) trained on a 20-nm mesh is evaluated on 12.5-nm and 9.375-nm meshes without retraining. \textbf{a}, One fixed, learned first-layer discrete–continuous (DISCO) kernel on its 60-nm-radius support (dashed circle), shown as a continuous function (left) and sampled at the grid points (dots) of the 20-, 12.5- and 9.375-nm meshes. \textbf{b}, One layout at normal incidence on the 20-, 12.5- and 9.375-nm meshes (rows). Columns show the ground truth from finite-difference time-domain simulation, DANO prediction and their error (DANO minus ground truth) for Re$\,E_{s,x}$. The blue frame marks the 20-nm training mesh. \textbf{c}, Permittivity of one pillar from \textbf{b} on the three meshes. Gray grid cells show fractional pillar coverage. \textbf{d}, Mean normalized mean squared error (NMSE) over 50 layouts at each mesh spacing.}
    \label{ed:invariants}
\end{figure}
\clearpage

\begin{figure}[h!]
    \centering
    \includegraphics[width=1.0\linewidth]{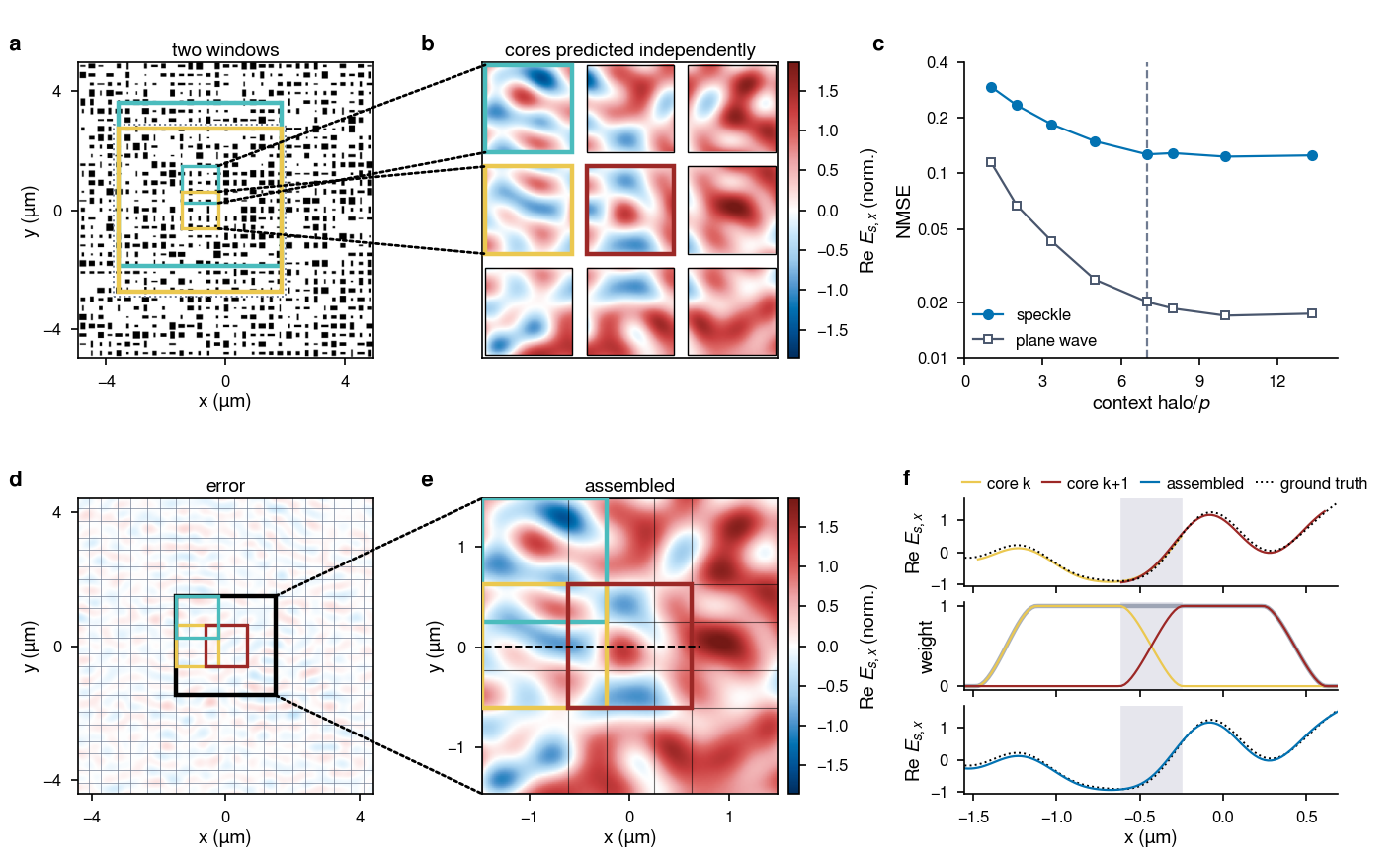}
    \caption{\textbf{Independent window predictions and partition-of-unity assembly.} \textbf{a}, Layout with two overlapping input windows (gold and teal). Inner squares mark the $1.24$-\textmu m prediction cores, and outer squares include a $2.1$-\textmu m context halo on each side. Dotted square, the permittivity crop of the gold window. \textbf{b}, Nine overlapping prediction cores, evaluated independently in their respective windows and separated for display before assembly. \textbf{c}, Validation core normalized mean squared error (NMSE) against context-halo width in constituent spacings, with one training run per width. Points are means over 600 speckle or 71 plane-wave cases. Vertical dashed line, the selected halo of 7 spacings (2.1\,\textmu m). \textbf{d}, Assembled-field error of the domain-decomposition attention neural operator (DANO), DANO minus the ground truth from finite-difference time-domain simulation. Grid lines mark core boundaries, and the black square marks the region enlarged in \textbf{e}. \textbf{e}, Assembled field in that region, with colored outlines marking overlapping prediction cores. Dashed horizontal line, the cut in \textbf{f}. \textbf{f}, Independent core predictions (top), partition-of-unity weights summing to one (middle) and the assembled field (bottom) along the cut. Shading, the overlap.}
    \label{ed:window}
\end{figure}
\clearpage

\begin{figure}[h!]
    \centering
    \includegraphics[width=1.0\linewidth]{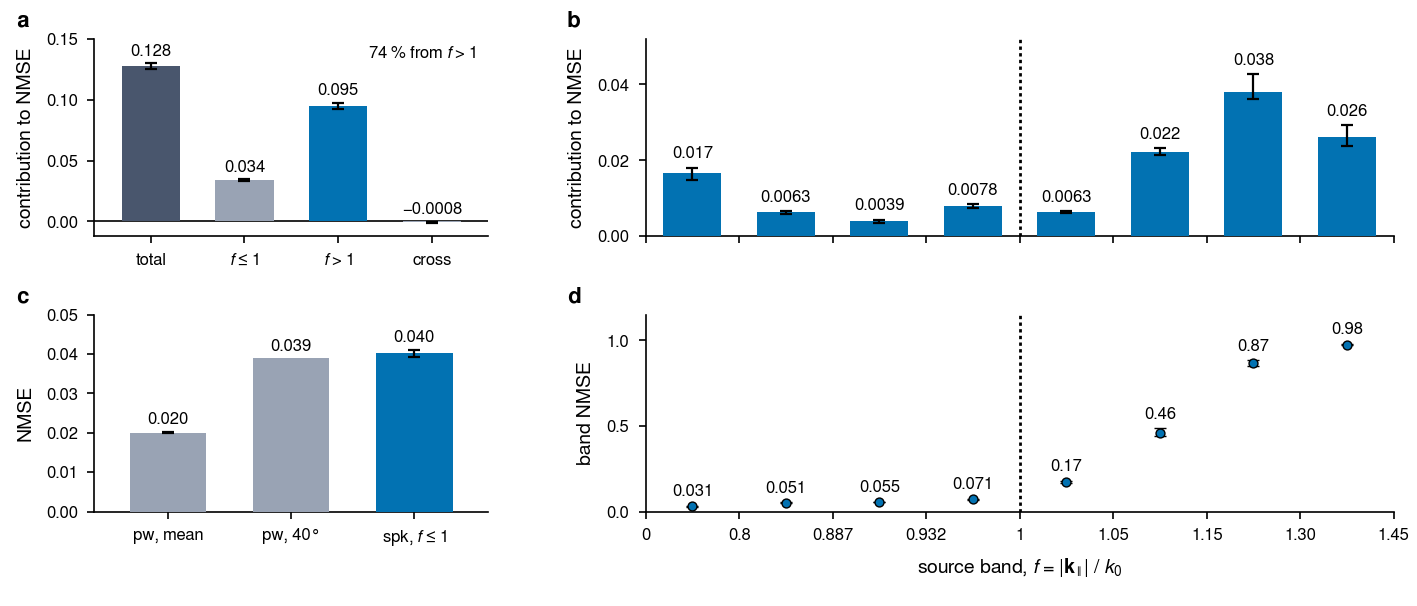}
    \caption{\textbf{Speckle test error by source band.} Errors of the domain-decomposition attention neural operator (DANO) on held-out speckle test cases. Source bands group the angular components of the speckle source by $f=|\mathbf{k}_\parallel|/k_0$, where $\mathbf{k}_\parallel$ is the in-plane wavevector and $k_0$ is the free-space wavenumber. \textbf{a}, Normalized mean squared error (NMSE) split into $f\le1$, $f>1$ and cross terms. \textbf{b}, Contribution of each of eight bands to the NMSE, normalized as in \textbf{a}, without cross terms between bands. \textbf{c}, NMSE for plane waves (pw), averaged over all incidence angles and at $40^\circ$, and for the $f\le1$ part of the speckle illumination (spk). \textbf{d}, NMSE of each band alone, normalized by the scattered-field energy of that band. Dotted lines in \textbf{b},\textbf{d}, $f=1$. Bands of unequal width are drawn at equal spacing. Error bars, 95\% bootstrap intervals over 32 realizations with two layouts each (\textbf{a} and spk in \textbf{c}) or over 213 cases (pw mean in \textbf{c}), and interquartile ranges over 12 realizations with one layout each (\textbf{b},\textbf{d}).}
    \label{ed:speckleband}
\end{figure}
\clearpage

\begin{figure}[h!]
    \centering
    \includegraphics[width=1.0\linewidth]{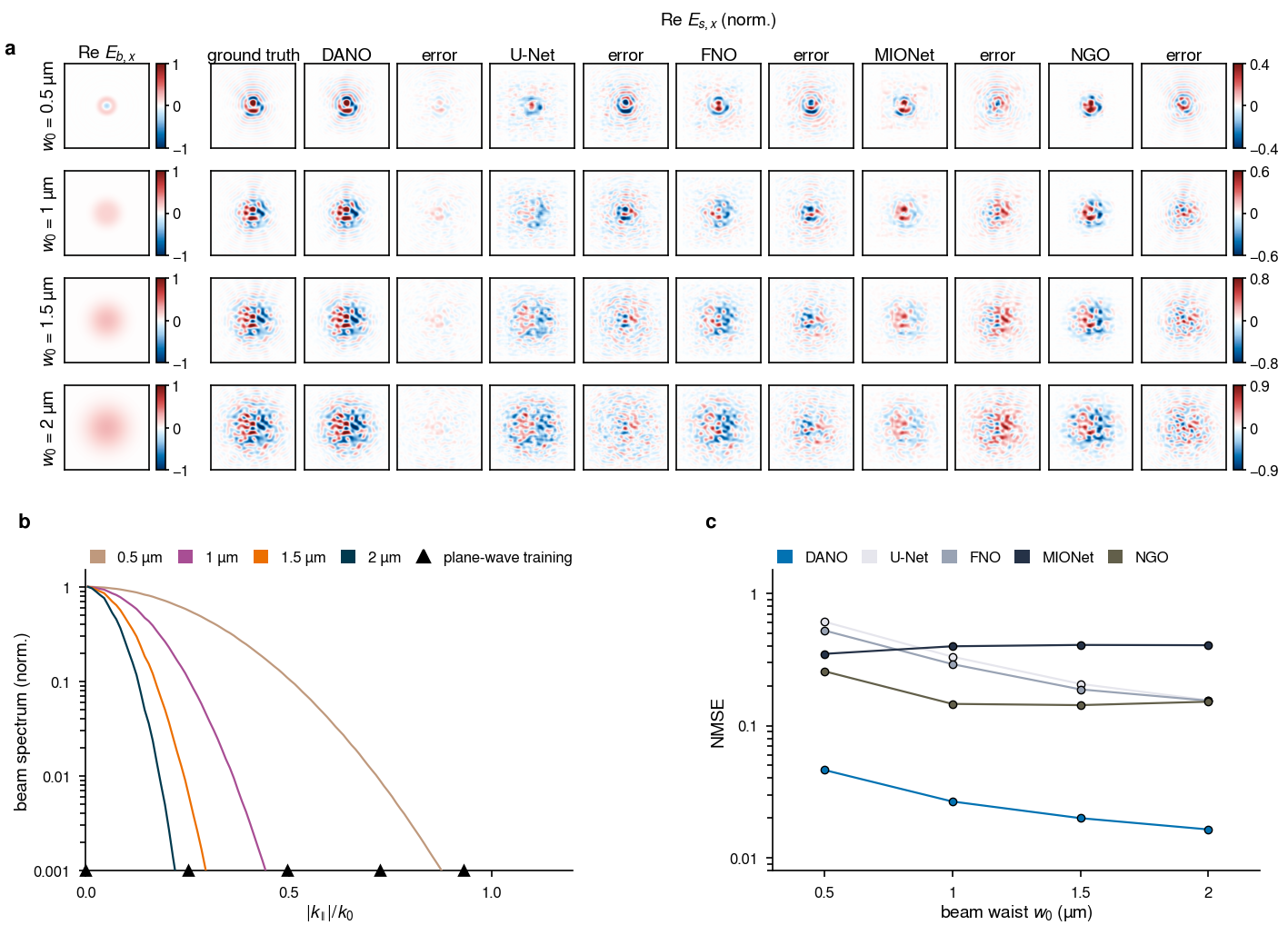}
    \caption{\textbf{Scattered-field prediction under Gaussian-beam illumination.} One held-out layout is illuminated by normally incident, $x$-polarized Gaussian beams with waists $w_0$ of 0.5, 1, 1.5 and 2\,\textmu m. The domain-decomposition attention neural operator (DANO), U-Net, Fourier neural operator (FNO), multiple-input operator network (MIONet) and neural Green's operator (NGO) were trained on the plane-wave and speckle corpora. \textbf{a}, Rows show beam waist. Columns show the background field $E_{b,x}$, the ground-truth scattered field $E_{s,x}$ from finite-difference time-domain simulation, and each model’s prediction and error (prediction minus ground truth). Real parts are normalized by the peak background-field amplitude in each row. \textbf{b}, Normalized beam angular spectra against in-plane wavevector. Black triangles mark the plane-wave training settings. \textbf{c}, Prediction-core normalized mean squared error (NMSE) against beam waist for DANO and the four baselines.}
    \label{ed:gaussian}
\end{figure}
\clearpage

\begin{figure}[h!]
    \centering
    \includegraphics[width=1.0\linewidth]{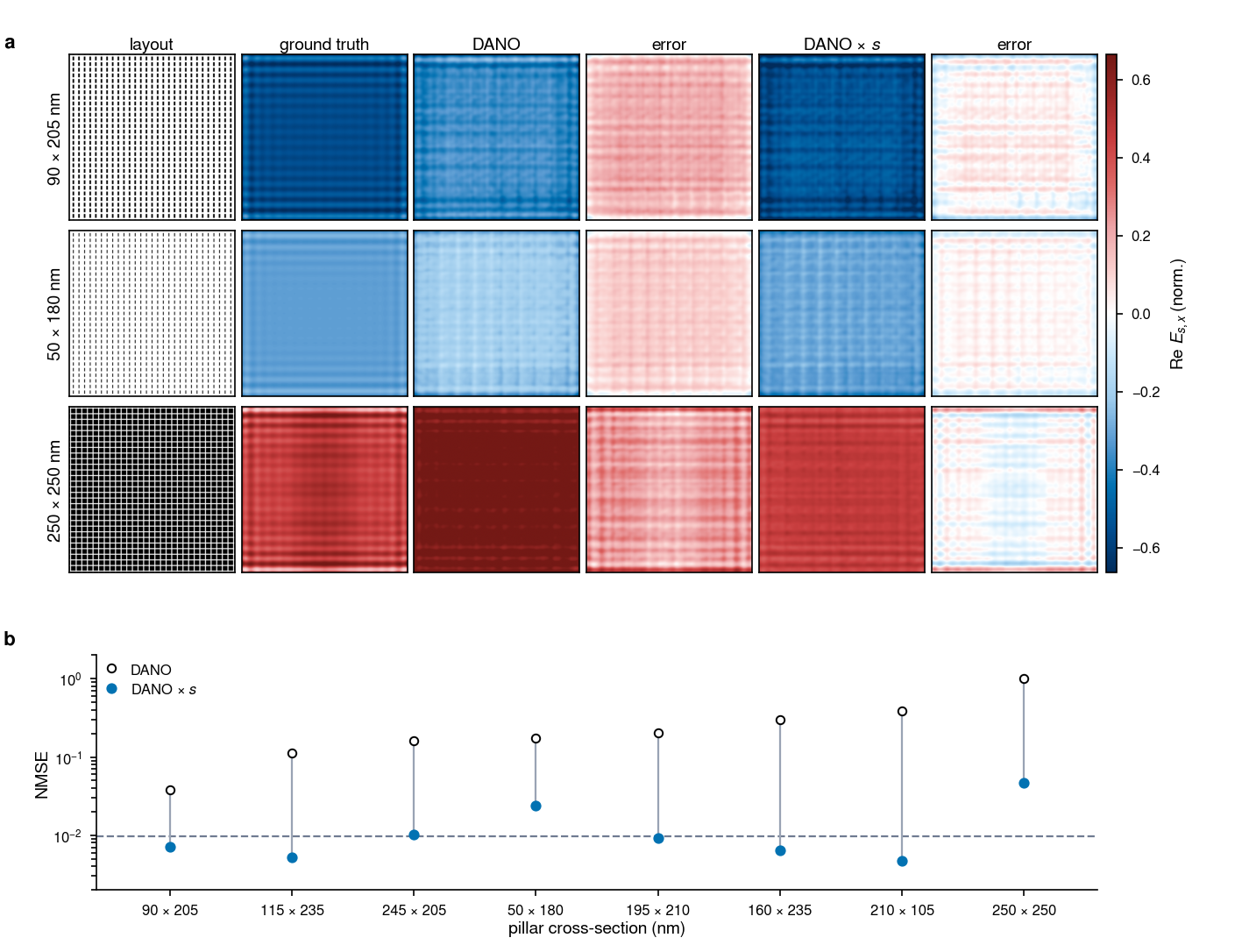}
    \caption{\textbf{Full-field prediction for uniform periodic arrays.} Eight fully occupied $33\times33$ arrays, each repeating one pillar cross-section on every site, are illuminated by a normally incident plane wave, with one array per cross-section. \textbf{a}, Arrays with the smallest, median and largest normalized mean squared error (NMSE) of the prediction (rows), with the cross-section at left. Columns show the layout, the ground truth from finite-difference time-domain simulation, the domain-decomposition attention neural operator (DANO) prediction and their error (DANO minus ground truth), followed by DANO $\times\,s$, the prediction multiplied by the single complex number $s$ that best fits the ground truth, and its error. All field maps show Re$\,E_{s,x}$, normalized by the root-mean-square incident-field amplitude on one color scale. \textbf{b}, NMSE of the assembled scattered field for the eight arrays, before (open circles) and after (filled circles) multiplication by $s$. Dashed line, mean NMSE of the 21 normal-incidence plane-wave test cases on random layouts.}
    \label{ed:ordered}
\end{figure}
\clearpage

\end{document}